\newcommand{\EPnum} {CERN-EP-2000-050}
\newcommand{\Date}      {06 Apr 2000}

\documentstyle[a4p,11pt,epsfig,array,cite,pennames,amssymb]{article}
\newcommand{\inmath}[1] {\ifmmode#1\else$#1$\fi}
\newcommand{\definmath}[2] {\def#1{\ifmmode#2\else$#2$\fi}}
\definmath{\PWpm} {\mathrm{W}^{\pm}}      
\definmath{\Pgtp} {\tau^{+}}        
\definmath{\Pgtm} {\tau^{-}}        
\definmath{\Pgtpm}   {\tau^{\pm}}         
\definmath{\Pgn}  {\nu}          
\definmath{\Pagn} {\overline{\nu}}     
\definmath{\Pq}      {\mathrm{q}}
\definmath{\Paq}  {\overline{\mathrm{q}}}
\definmath{\Pu}      {\mathrm{u}}
\definmath{\Pau}  {\overline{\mathrm{u}}}
\definmath{\Pd}      {\mathrm{d}}
\definmath{\Pad}  {\overline{\mathrm{d}}}
\definmath{\Ps}      {\mathrm{s}}
\definmath{\Pas}  {\overline{\mathrm{s}}}
\definmath{\Pc}      {\mathrm{c}}
\definmath{\Pac}  {\overline{\mathrm{c}}}
\definmath{\Pb}      {\mathrm{b}}
\definmath{\Pab}  {\overline{\mathrm{b}}}
\definmath{\Pt}      {\mathrm{t}}
\definmath{\Pat}  {\overline{\mathrm{t}}}
\definmath{\Pap}  {\overline{\mathrm{p}}}
\definmath{\Pan}  {\overline{\mathrm{n}}}
\definmath{\PaD}  {\overline{\mathrm{D}}}
\definmath{\PaDz} {\overline{\mathrm{D}}^{0}}
\definmath{\PaB}  {\overline{\mathrm{B}}}
\definmath{\PaBz} {\overline{\mathrm{B}}^{0}}
\definmath{\PsDpm}   {\mathrm{D}^{\pm}_{\mathrm{s}}}  
\definmath{\PcgLpm}  {\Lambda^{\pm}_{\mathrm{c}}}  
\definmath{\PD} {\mathrm{D}}     
\definmath{\PDst} {\mathrm{D}^{*}}     
\definmath{\PgLz} {\Lambda^{0}}        
\newcommand {\lsp}      {{{\tilde{\chi}}^{0}}_{1}}
\newcommand {\nln}      {{{\tilde{\chi}}^{0}}_{2}}
\newcommand {\Gravitino} {\tilde{\mathrm{G}}}
\newcommand {\gv} {\tilde{\mathrm{G}}}
\newcommand {\nngg}       {\nu\overline{\nu}\gamma\gamma}
\newcommand {\nnggbra}       {\nu\overline{\nu}\gamma(\gamma)}
\newcommand {\nngggbra}       {\nu\overline{\nu}\gamma\gamma(\gamma)}
\newcommand {\ra}         {\rightarrow}
\newcommand {\ee}         {\mathrm{e}^+ \mathrm{e}^-}

\newcommand{\epem}   {\Pep\Pem}
\newcommand{\gamgam} {\Pgg\Pgg}
\newcommand{\mumu}   {\Pgmp\Pgmm}
\newcommand{\tautau} {\Pgtp\Pgtm}
\newcommand{\nunu}   {\Pgn\Pagn}

\newcommand{\eetogg}    {\epem\to\gamgam}

\newcommand{\eetomumu}     {\epem\to\mumu}
\newcommand{\eetotautau}   {\epem\to\tautau}

\newcommand{\eetonunu}     {\epem\to\nunu}

\newcommand{\BR}             {{\mathrm BR}}
\newcommand{\PX}             {{\mathrm X}}
\newcommand{\PY}             {{\mathrm{Y}}}
\newcommand{\eetonngg}       {\epem \to \nngg}
\newcommand{\eetonnggbra}    {\epem \to \nnggbra}
\newcommand{\eetonngggbra}   {\epem \to \nngggbra}
\newcommand{\eetoXY}         {\epem \to \PX\PY}
\newcommand{\eetoXX}     {\epem \to \PX\PX}
\newcommand{\XtoYg}       {\PX \to \PY\gamma}

\newcommand{\sigbrXY}    {\sigma(\eetoXY)\cdot\BR(\XtoYg)} 
\newcommand{\sigbrXX}    {\sigma(\eetoXX)\cdot\BR^2(\XtoYg)} 
\newcommand{\eetoXXs}   {\epem \to \nln \nln}
\newcommand{\eetoXYs}   {\epem \to \nln \lsp}
\newcommand{\XtoYgs}    {\nln \to \lsp\gamma}

\newcommand{\mx}         {M_{\PX}}
\newcommand{\my}         {M_{\PY}}
\newcommand{\myzero}        {\my\approx 0}

\newcommand{\mxmax}      {\mx^{\rm max}}

\newcommand{\roots} {\sqrt{s}}

\newcommand{\costhe} {\cos\theta}
\newcommand{\acosthe}   {|\costhe\,|}

\definmath{\GeV}  {\mathrm{GeV}}
\definmath{\GeVc} {\mathrm{GeV}\!/c}
\definmath{\GeVcc}   {\mathrm{GeV}\!/c^2}
\definmath{\MeV}  {\mathrm{MeV}}
\definmath{\MeVc} {\mathrm{MeV}\!/c}
\definmath{\MeVcc}   {\mathrm{MeV}\!/c^2}
\definmath{\MVm}  {\mathrm{MV}\!/\mathrm{m}}
\definmath{\keV}  {\mathrm{keV}}
\definmath{\keVcm}   {\mathrm{keV}\!/\mathrm{cm}}
\definmath{\kV}      {\mathrm{kV}}
\definmath{\km}      {\mathrm{km}}
\definmath{\meter}   {\mathrm{m}}
\definmath{\cm}      {\mathrm{cm}}
\definmath{\mm}      {\mathrm{mm}}
\definmath{\micron}  {\mu\mathrm{m}}
\definmath{\nm}      {\mathrm{nm}}
\definmath{\kg}      {\mathrm{kg}}
\definmath{\gram} {\mathrm{g}}
\definmath{\second}  {\mathrm{s}}
\definmath{\microsec}   {\mu\mathrm{s}}
\definmath{\degree}  {^\circ}
\definmath{\degC} {^\circ\mathrm{C}}
\definmath{\ohm}  {\Omega}
\definmath{\Mohm} {\mathrm{M}\Omega}
\definmath{\rad}  {\mathrm{rad}}
\definmath{\mrad} {\mathrm{mrad}}
\definmath{\nb}      {\mathrm{nb}}

\newcommand{\eqref}[1]  {(\ref{#1})}
\newcommand{\NIM} {Nucl.~Instrum.\ Methods}

\newcommand{\OPALColl}  {OPAL Collab.}

\newcolumntype{L} {>{$}l<{$}}
\newcolumntype{C} {>{$}c<{$}}
\newcolumntype{R} {>{$}r<{$}}

\newcommand{\grcnng} {\texttt{grc$\nu\nu\gamma$}}

\begin{document}
\begin{titlepage}
\begin{center}{\large   EUROPEAN ORGANIZATION FOR NUCLEAR RESEARCH
}\end{center}\bigskip
\begin{flushright}
  \EPnum\\
  \Date \\
\end{flushright}

\begin{center}{\huge\bf \boldmath 
  Photonic Events with Missing Energy \\
  in $\epem$ Collisions at $\roots$ = 189~$\GeV$
}\end{center}\bigskip\bigskip
\begin{center}{\LARGE The OPAL Collaboration
}\end{center}\bigskip \bigskip

\begin{abstract} 
Photonic events with large missing energy have been observed in 
$\mathrm e^+e^-$ collisions at a centre-of-mass energy of 189~GeV
using the OPAL detector at LEP. Results are presented for event 
topologies consistent with a single photon or with an acoplanar 
photon pair.
Cross-section measurements are performed within the kinematic acceptance
of each selection, and the number of light neutrino species is
measured.  Cross-section results are compared with the expectations from
the Standard Model process $\eetonunu$ + photon(s). No evidence is
observed for new physics contributions to these final states. 
Upper limits on $\sigbrXY$ and $\sigbrXX$ are derived
for the case of stable and invisible $\PY$.  These limits apply to
single and pair production of excited neutrinos ($\PX = \nu^*, \PY =
\nu$), to neutralino production ($\PX=\nln, \PY=\lsp$) and to
supersymmetric models in which $\PX = \lsp$ and $\PY=\Gravitino$ is a
light gravitino.  The case of 
macroscopic decay lengths of particle X 
is considered
for $\rm \epem \to XX$, $\rm X \to Y \gamma$, when
$\myzero$.  The single-photon results are also used to place upper
limits on superlight gravitino pair production as well as 
graviton-photon production in the context of theories with
additional space dimensions.

\end{abstract}

\bigskip\bigskip
\begin{center}{\large
  Submitted to Eur. Phys. J. {\bf C}
}\end{center}


\end{titlepage}

%

\begin{center}{\Large        The OPAL Collaboration
}\end{center}\bigskip
\begin{center}{
G.\thinspace Abbiendi$^{  2}$,
K.\thinspace Ackerstaff$^{  8}$,
C.\thinspace Ainsley$^{  5}$,
P.F.\thinspace Akesson$^{  3}$,
G.\thinspace Alexander$^{ 22}$,
J.\thinspace Allison$^{ 16}$,
K.J.\thinspace Anderson$^{  9}$,
S.\thinspace Arcelli$^{ 17}$,
S.\thinspace Asai$^{ 23}$,
S.F.\thinspace Ashby$^{  1}$
% ...
}\end{center}\bigskip
\bigskip
$^{  1}$School of Physics and Astronomy, University of Birmingham,
Birmingham B15 2TT, UK
\newline
$^{  2}$Dipartimento di Fisica dell' Universit\`a di Bologna and INFN,
I-40126 Bologna, Italy
\newline
$^{  3}$Physikalisches Institut, Universit\"at Bonn,
D-53115 Bonn, Germany
\newline
% ...
$^{  5}$Cavendish Laboratory, Cambridge CB3 0HE, UK
\newline
% ...
$^{  8}$CERN, European Organisation for Nuclear Research,
CH-1211 Geneva 23, Switzerland
\newline
$^{  9}$Enrico Fermi Institute and Department of Physics,
University of Chicago, Chicago IL 60637, USA
\newline
% ...
$^{ 16}$Department of Physics, Schuster Laboratory, The University,
Manchester M13 9PL, UK
\newline
$^{ 17}$Department of Physics, University of Maryland,
College Park, MD 20742, USA
\newline
% ...
$^{ 22}$Department of Physics and Astronomy, Tel Aviv University,
Tel Aviv 69978, Israel
\newline
$^{ 23}$International Centre for Elementary Particle Physics and
Department of Physics, University of Tokyo, Tokyo 113-0033, and
Kobe University, Kobe 657-8501, Japan
\newline
% ...
Budapest, Hungary.

\clearpage\newpage
\section{ Introduction }
\label{sec:intro}
 
We describe measurements and searches using
a data sample of
photonic events with large missing
energy collected in 1998 with the OPAL detector at LEP.
The events result
from $\epem$ collisions at a centre-of-mass energy 
of 188.6~GeV with an integrated luminosity of 
177.3~pb$^{-1}$. 
The present paper builds on publications 
from earlier data samples at lower centre-of-mass energies
\cite{OPALSP183,OPALSP172}.
This      
data-set at 189 GeV
gives discovery potential in a new kinematic regime
with about a four-fold increase in integrated luminosity.
Measurements of 
photonic event production have also been made by the other LEP
collaborations at centre-of-mass energies above the W pair 
threshold~\cite{LEP2SP}, including new results 
from L3 and DELPHI at $\sqrt{s}=$ 
189~GeV~\cite{LEPSP189}.

The single-photon  and acoplanar-photons search topologies presented here
are designed to select events with one or more photons and significant missing 
transverse energy, indicating  the presence of at least one neutrino-like 
invisible particle which interacts only weakly with matter.
The event selections for these search topologies are 
similar to those used in our recent publication\cite{OPALSP183}.
The single-photon search topology is sensitive to events in which 
there are one or two photons and missing energy which, within the 
Standard Model, are expected from the $\eetonnggbra$ 
process\footnote{The photon in parentheses denotes that the
presence of this photon is allowed but not required.}.
The acoplanar-photons search topology is designed to select events with 
two or more photons and significant missing transverse energy
which, within the Standard Model, are expected 
from the $\eetonngggbra$ process. 

The single photon topology provides a 
direct measurement of the invisible width of the $\mathrm{Z}^{0}$ 
and can probe charged and neutral triple gauge couplings.
The acoplanar-photons topology can probe WW$\gamma \gamma$ quartic 
couplings in
the 
$\ee \ra \nu_{\mathrm{e}} \overline{\nu_{\mathrm{e}}} \gamma \gamma$ process.
The neutral and quartic gauge coupling measurements
will be described in forthcoming papers based on
the event selections described herein.

These photonic final-state topologies are sensitive to several different 
new physics scenarios. 
A generic classification is
$\eetoXY$ or $\eetoXX$ where $\PX$ is neutral and can decay 
radiatively 
($\XtoYg$) and $\PY$ is stable and only weakly interacting. For the general case of
massive $\PX$ and $\PY$ this includes conventional supersymmetric
processes $(\PX = \nln, \PY = \lsp)$. 
These topologies also have particularly good sensitivity
for the special case of $\myzero$.
This applies both to the production of
excited neutrinos $(\PX = \nu^*, \PY = \nu)$ and to supersymmetric models
in which the lightest supersymmetric particle (LSP) is a light 
gravitino
and $\lsp$ is the next-to-lightest supersymmetric particle (NLSP) which decays 
to a gravitino and a photon ($\PX=\lsp, \PY=\Gravitino$).
The neutralino life-time in such models is a free parameter and so we
also address the possibility of neutralino-pair production 
with macroscopic decay lengths.
One type of new physics 
which could be seen in the single-photon topology 
is the production of an invisible particle in association with a photon.
An example of this is
graviton-photon production, $\epem \ra \mathrm{G} \gamma$~\cite{GRW,Peskin}.
This can occur within string theory models
which allow gravitons to propagate in a higher-dimensional space
but restrict Standard Model particles to the
usual four space-time dimensions~\cite{ADD}.
Another type of new physics is
the production of invisible particles tagged by initial-state radiation.
One example\footnote{The initial-state radiation diagram is one of
many that contribute
to this final state.}
is production of a pair of gravitinos,
$\epem \ra \Gravitino \Gravitino \gamma$, as in the
superlight gravitino model~\cite{Zwirner}.
The acoplanar-photons search topology also has sensitivity to the
production of two particles, one invisible, or with an invisible decay mode, 
and the other decaying into two photons. 

This paper will first describe the OPAL detector
and the Monte Carlo samples used.
A brief summary of the
event selections will
then be given,
followed by cross-section measurements for 
$\eetonnggbra$ and $\eetonngggbra$ and comparisons with Standard Model 
expectations.
The new physics search results 
will then be discussed. 

\section{Detector and Monte Carlo Samples}
\label{sec:detector}

The OPAL detector, which is described in detail in~\cite{OPAL-detector},
contains a silicon micro-vertex detector surrounded by
a pressurized
central tracking system operating inside
a solenoid with a magnetic field of 0.435 T.
The region
outside the 
solenoid (barrel) and the pressure bell (endcap) is 
instrumented with scintillation counters, presamplers and the 
lead-glass electromagnetic calorimeter (ECAL).
The magnet return yoke is instrumented for hadron calorimetry and
is surrounded by external muon chambers.
Electromagnetic calorimeters close to the beam
axis
measure luminosity and complete
the acceptance.

The measurements presented here are mainly based on the observation of 
clusters of energy deposited in the lead-glass electromagnetic 
calorimeter. This consists of an array of 9,440 lead-glass blocks
in the barrel ($|\cos{\theta}| < 0.82$) with a
quasi-pointing geometry and two dome-shaped endcap arrays, each of 1,132
lead-glass blocks,
covering the polar angle\footnote{In the OPAL coordinate system, 
$\theta$ is the polar angle defined with respect to the electron 
beam direction and $\phi$ is the azimuthal angle.} 
range ($0.81 < |\cos{\theta}| < 0.984$).
Fully hermetic electromagnetic calorimeter coverage is achieved 
beyond the end of the ECAL down to small polar angles with the use of the
the gamma-catcher calorimeter, the forward 
calorimeter (FD) and the silicon-tungsten calorimeter (SW). 

Scintillators in the barrel and endcap 
regions are used to 
reject backgrounds from cosmic ray 
interactions by providing time measurements for 
the large fraction ($\approx$ 80\%) of
photons which convert in the material in front of the ECAL.
The barrel time-of-flight (TOF) scintillator bars
are located outside the solenoid 
in front of the barrel ECAL
and match  
its geometrical 
acceptance ($|\cos{\theta}| < 0.82$).
Tile endcap (TE) scintillator arrays~\cite{OPAL-TE} are 
located at $0.81 < \acosthe < 0.955$
behind the pressure bell and 
in front of the endcap ECAL. 


The integrated luminosities of the data samples 
are determined to better than 1\% from small-angle Bhabha
scattering events in the SW calorimeter.
Triggers\cite{trigger} based on electromagnetic energy deposits in 
either the barrel or endcap electromagnetic calorimeters
lead to full trigger efficiency for photonic events passing the 
event selection criteria described in the following section.
 
The 
KORALZ~\cite{KORALZ} and NUNUGPV98~\cite{NUNUGPV98}
Monte Carlo generators
were used to simulate the 
expected Standard Model signal process,  
$\eetonunu$ + photon(s).
For other expected Standard Model processes, a number of different 
generators were used: RADCOR~\cite{RADCOR} for 
$\epem \to \gamma \gamma (\gamma)$;
BHWIDE~\cite{BHWIDE} for $\epem \to \epem$;
TEEGG~\cite{TEEGG} for $\epem \to \epem \gamma$;
KORALW~\cite{KORALW} using \texttt{grc4f}~\cite{GRC4F} matrix elements for
$\ee \to \ell^+ \ell^- \nu \bar{\nu} (\gamma)$ 
and
$\ee \to \nu \bar{\nu} \mathrm{q} \bar{\mathrm{q}} $, 
and KORALZ for  $\eetomumu(\gamma)$ and $\eetotautau(\gamma)$.
The Vermaseren program\cite{VERMASEREN} and \texttt{grc4f} were used
for $\ee \to \ee \ell^+ \ell^-$.
The expected contributions from each of these Standard 
Model processes were evaluated
using a total equivalent integrated luminosity at least five
times larger than the integrated 
luminosity of the data sample.

To simulate possible new physics processes of the type  $\eetoXY$ 
and $\eetoXX$ where $\PX$ decays to $\PY\gamma$ and 
$\PY$ escapes detection, a modified version of
the SUSYGEN~\cite{SUSYGEN} Monte Carlo 
generator was used to produce neutralino pair events of the type 
$\eetoXYs$ and  $\eetoXXs$, $\XtoYgs$,  
with isotropic angular
distributions for the production and decay of $\nln$ and 
including initial-state radiation.
Monte Carlo events were generated at 48 (for $\PX\PY$ production) and 42
(for $\PX\PX$ production) points in the kinematically accessible region of the 
($\mx$, $\my$) plane.
A Monte Carlo generator was written to simulate 
the superlight gravitino
signature $\gv \gv \gamma$, discussed in Section~\ref{sec:sp_results_GGg}.
The KORALZ $\nnggbra$ sample was used to determine the efficiency for 
graviton-photon production in the context of additional space dimensions, 
by means of an event reweighting (see section~\ref{sec:sp_results_Gg}).
The same procedure was also used to calculate the efficiency for
$\gv \gv \gamma$ production, and compared with the prediction of the 
direct Monte Carlo simulation.
All the Monte Carlo samples described above were processed through the
OPAL detector simulation~\cite{GOPAL}.

Simulation of  $\eetoXX$, $\XtoYg$ signal events with $\myzero$ where 
$\mx$ has a finite non-zero lifetime $\tau_X$ was 
implemented in
the full simulation of the OPAL detector. In particular,
the massive
quasi-stable neutral particle $\PX$ was assigned properties similar to
a heavy neutrino for the purpose of propagation, and then propagated
within the GEANT~\cite{ref:geant} framework, according to its initial
kinematics and $\tau_X$ before forcing the $\XtoYg$ decay.

\section{Photonic Event Selection }
\label{sec:selection}

This section summarizes the criteria for selecting single-photon and 
acoplanar-photons events.
The kinematic acceptance of each selection is defined in terms of the photon energy, 
$E_{\gamma}$, and the photon polar angle, $\theta$. In addition,  
the scaled energy, $x_{\gamma}$, is defined as
$E_{\gamma}/E_{\mathrm{beam}}$, and the scaled transverse
energy, $x_{T}$, as $x_{\gamma}\sin{\theta}$.

\begin{description}
\item[Single-Photon -]
One or two photons accompanied by invisible particle(s):
\begin{itemize}
\item At least one photon with  $x_{T} > 0.05$ and with
$15^{\circ}<\theta<165^{\circ}$ ($\acosthe < 0.966)$.
\end{itemize}
\item[Acoplanar-Photons -] 
Two or more photons accompanied by invisible particle(s):
\begin{itemize}
\item At least two photons, each with $x_{\gamma} > 0.05 $
and $15^{\circ}<\theta<165^{\circ}$, or one photon 
with $E_{\gamma} > 1.75$ GeV  and $\acosthe < 0.8$ and a second photon
with $E_{\gamma} > 1.75$ GeV and $15^{\circ}<\theta<165^{\circ}$.
\item The transverse momentum $p_T^{\gamma\gamma}$
of the two-photon system 
consisting of the two highest energy photons must satisfy 
$p_T^{\gamma\gamma}/E_{\rm beam} > 0.05$.
\end{itemize}
\end{description}

In each of the two cases we retain acceptance for events 
with additional photons in which the resulting photonic system is
still consistent with the presence of significant 
missing energy. This reduces the sensitivity of each measurement to the
modelling of higher-order contributions.
Consequently,
a large fraction of the kinematic acceptance of the acoplanar-photons
selection is also contained in the kinematic acceptance of the
single-photon 
selection.

\subsection{Single-Photon Event Selection}
\label{sec:g1_selection}

The single-photon selection criteria are explained in detail in
a previous publication\cite{OPALSP183}. 
After defining the kinematic acceptance,
additional cuts on cluster quality, forward energy, muon chamber
and hadron calorimeter information, and multiphoton kinematics
are used to remove cosmic ray backgrounds, beam related backgrounds
and standard model physics backgrounds that have no missing
energy. Then events are classed as either having or not having
photon conversion candidates based on tracking chamber information.
Each of these two classes has somewhat different additional
selection criteria based on charged track activity, time-of-flight
information, and other background suppression cuts. 

Several improvements have been made to 
the single-photon selection used 
in this analysis compared with the previous analysis\cite{OPALSP183}.
They are described in detail in Appendix A.

\subsection{Acoplanar-Photons Event Selection}
\label{sec:g2_selection}

The acoplanar-photons selection has classes of selection
requirements similar to those of the single-photon selection. The details 
of the acoplanar-photons selection are described in our previous
analysis\cite{OPALSP183}; 
changes with respect to the previous analysis are minimal and are 
given in Appendix A.


The requirement of a second photon in an event reduces many of the 
backgrounds which are otherwise a problem for the single-photon analysis. 
For this reason, the acoplanar photons selection can tolerate a
lower energy threshold for the most energetic photon, as well as
looser but more inclusive acceptance requirements for photon
conversions. On the other hand, the single-photon selection
has more acceptance for events having no time-of-flight information
for the photons. 
In order to obtain the best overall acceptance for acoplanar-photons, we 
have added to the acoplanar-photons selection
that part of the single-photon selection which 
contains two photons
within the 
kinematic acceptance of the acoplanar-photons selection.
This addition results in a relative increase in
efficiency of 9.6\% for Standard Model $\eetonngggbra$ events.

\section{Selection Results}
\label{sec:results}

The results of the single-photon and acoplanar-photons selections
and the corresponding cross-section measurements and 
other measured event quantities are given below in 
sections~\ref{sec:sp_results} and~\ref{sec:g2_results}.
Results from both selections are summarized in 
Table~\ref{tab:sp_results}.

\boldmath
\subsection{Single-Photon}
\unboldmath
\label{sec:sp_results}

After applying the single-photon selection criteria 
to the data sample, 643 events are selected. 
The expected contribution from cosmic ray and beam-related 
backgrounds is 4.6 $\pm$ 1.5 events.
These backgrounds have been estimated from events
having out-of-time TOF or TE information, but passing all other selection
criteria, and from events selected with looser criteria that have 
been visually scanned. 
Of the expected physics backgrounds from plausible sources,
$\ee \ra \ee \gamma$,
$\ee \ra \ell^+ \ell^- \nu\bar{\nu} (\gamma)$, 
$\rm \ee \ra \ee \ell^+ \ell^- $,
$\rm \ee \ra \nu \bar{\nu} q\bar{q}$,
$\ee \ra \mumu \gamma$ and
$\ee \ra \tautau \gamma$ have non-negligible 
contributions\footnote{
The expected number of events from the Standard Model process 
$\ee \ra \nu \bar{\nu} \nu\bar{\nu} \gamma (\gamma)$ is negligible and is
neglected in both the efficiency and background estimates}. 
The total number of expected physics
background events is 4.6 $\pm$ 0.5 and the 
contributing sources 
are summarized in Table~\ref{tab:sp_background}.
%
The number of events expected from the Standard Model process
$\eetonunu \gamma(\gamma)$ as predicted by KORALZ is 
679 $\pm$ 5 (stat) $\pm$ 14 (sys) where the systematic error is from
experimental sources as discussed later.
The number of events observed
is consistent
with the number expected from 
$\eetonnggbra$ plus the background.
The efficiency for selecting $\eetonnggbra$ events within the
kinematic acceptance of the single-photon selection
is $(82.1 \pm 1.7)$\%.
For both the single-photon and acoplanar-photons selections,
efficiency losses due to vetoes on random
detector occupancy range from about 
(3-5)\%.  Quoted efficiencies include these losses.
The single-photon cross-section for $\sqrt{s}=189$~GeV, 
accounting for detector and 
selection efficiencies and subtracting the estimated background,
is 4.35 $\pm$ 0.17 (stat) $\pm$ 0.09 (sys)~pb.

The systematic error on the cross-section measurement
from knowledge of the efficiency and normalisation 
is estimated to be 2.1\%.
The contributing uncertainties are summarized in 
Table~3.
The dominant systematics are 
the selection efficiency uncertainty (1.5\%) and 
the uncertainty on the detector occupancy estimate (1\%).
The event selection efficiency  
is controlled using a data sample of around 1200 
$\ee \ra \gamma \gamma$
events. 
In particular, the efficiency and time response of the TB and TE
scintillators are measured from these data and small correction 
factors applied to the efficiency. 
The selection efficiency systematic includes the
uncertainties on such corrections and an estimate of the 
residual uncertainty on the efficiency. 
The occupancy, estimated from random
beam-crossing triggers, is typically 4\% and we
assign an error of 1\%.
We have compared the estimated efficiency using 
the NUNUGPV98 and KORALZ event generators. We find that the
relative efficiency difference is ($0.5 \pm 0.4$)\% and
assign an error of 0.5\% to take into account the
sensitivity to the modelling of the 
photon energy, angle and multiplicity distributions.
This source
of error is considerably reduced from our 
previous publications, largely because both generators that are now
used are designed to model accurately $\nu \bar{\nu} \gamma \gamma$ events.
Additional systematic 
errors on the efficiency arise from uncertainties in 
modelling the material close to or inside the beam-pipe which
accounts for photons which convert early (20\% relative uncertainty)
and from systematic effects in the track reconstruction 
and parameter estimation which affect the association of tracks
to ECAL clusters.
Uncertainties in the photon energy scale, resolution and 
angular measurement give small
additional systematic contributions.

The cross-section as a function of centre-of-mass
energy is plotted in Figure~\ref{f:sp_xs130_189}.
Cross-section results from the current analysis and from our earlier 
publications\cite{OPALSP183,OPALSP172} 
are plotted.  The curve shows the
predicted cross-section from the KORALZ event generator
for the Standard Model process $\eetonnggbra$. 
The data are consistent with the prediction.

Figure~\ref{f:sp_data189}a shows the recoil mass distribution,
where the recoil mass $M_{\rm recoil}$ is defined as
the mass recoiling against the photon (or against the two-photon system).
The peak in the distribution at $M_{\rm Z}$ is due to a large 
contribution from the decay $\PZz \to \nunu$.
Figure~\ref{f:sp_data189}b shows the 
polar angle distribution along with the $\eetonnggbra$ 
Monte Carlo expectation. 
In both distributions, there is consistency between 
data and Monte Carlo, although we note a deficit in the 
radiative return peak and a slight excess in the low energy, high
recoil mass region.

The single-photon selection is designed to allow for the presence
of a second photon in order to accept events from the $\eetonngg$
process. Thirty-six observed events are considered to
be two-photon events (i.e.\ have a second photon with deposited
energy exceeding 300~MeV and with $15^{\circ}<\theta<165^{\circ}$),
consistent with the expectation  of 33.6 $\pm$ 1.5 
events from the KORALZ Monte Carlo.  Of these,
20 fall within the kinematic acceptance of the acoplanar photon
selection, as compared to the KORALZ expectation of 
$23.5 \pm 1.0$.

\boldmath
\subsection {Acoplanar-Photons}
\unboldmath
\label{sec:g2_results}

The acoplanar-photons selection applied to the data
sample yields 24 events, in good agreement with the KORALZ prediction 
of $26.9 \pm 1.2$ events for the Standard Model $\eetonngggbra$ contribution. 
The expected contribution from other Standard Model processes and 
from cosmic ray and beam related backgrounds is $0.11\pm 0.04$ events.
The selection efficiency for $\eetonngggbra$ events within the kinematic 
acceptance of the selection is 
($66.4\pm 2.9$)\%. 
The corresponding cross-section
is $0.204 \pm 0.043$~pb, 
compared to a KORALZ 
prediction of $0.228 \pm 0.002$~pb. 
The OPAL measurements of the cross-sections at $\roots =$ 130, 136, 161,
172, 183 and 189 GeV are summarized in Table~\ref{tab:g2_xsec_189}. 
Results for $\roots < 189$ GeV have been taken from our
previous publications\cite{OPALSP172,OPALSP183}.

The dominant source of systematic uncertainties is modelling of 
the reconstruction efficiency, especially the simulation of the detector 
material and consequent photon conversion probabilities. Other sources arise
from uncertainties on the electromagnetic calorimeter energy scale and 
resolution, on the integrated luminosity measurement, on  detector 
occupancy estimates and from comparisons of different Monte Carlo event 
generators for the process $\eetonngggbra$. The relative systematic error 
from all sources is 4.3\%. 

The kinematic properties of the selected events are displayed in
Figure~\ref{f:g2_kine_189} where they are compared with the predicted
distributions for $\eetonngggbra$ obtained using the KORALZ generator
normalized to the integrated luminosity of the data. Plot
(a) shows the recoil mass distribution of the selected acoplanar-photon
pairs (or of the two most-energetic photons in the case of events with 
three or more photons). 
The distribution is  peaked near the mass of the $\PZz$ as is 
expected for
contributions from $\eetonngggbra$.  The resolution of the recoil mass is
typically 3-5 GeV for $M_{\rm recoil}\approx M_{\rm Z}$.  Plot (b) shows
the distribution of the scaled energy of the second most energetic photon.
Plot (c) shows the $\gamma\gamma$ invariant-mass distribution for which
the mass resolution is typically 0.7-1.9 GeV. Plot (d) shows the
distribution in scaled transverse momentum of the selected two-photon
system. There is 1 selected event having a third photon with deposited energy above 
300 MeV and within the polar-angle acceptance of the selection. The corresponding
expectation  from KORALZ is $1.20 \pm 0.08$ events.

\section{Data Interpretation}
\label{sec:interpret}

The results of the single-photon and acoplanar-photons selections
are used to test the Standard Model and 
search for new physics contributions. 
%

For the XY and XX searches, we test the range
of the following product branching ratios 
which are consistent with the data:
$\rm \sigma ( {\mathrm e^+e^-\rightarrow XY} ) 
\cdot BR ( {\mathrm X\rightarrow Y\gamma} )$
and 
\mbox{$\rm \sigma ( {\mathrm e^+e^-\rightarrow XX} ) 
\cdot BR^2 ( {\mathrm X\rightarrow Y\gamma} )$}.
Given lack of evidence for signal,
we then set 95\% CL upper limits on these quantities.
This is done both for the general case of massive $\PX$ and $\PY$, 
and also separately for the special case of $\myzero$. 
All efficiencies are first evaluated under the assumption that the
decay length of $\PX$ is zero. 
For the XX search in the special case of $\myzero$, we
evaluate the efficiency as a 
function of the non-zero lifetime of the X particle, thus 
quantifying the sensitivity of the 
search for the general case of non-prompt decay.
For both the XY and XX searches, Monte Carlo samples
were generated for a variety of mass 
points in the kinematically accessible region of the $(\mx,\my)$ plane.
To set limits for arbitrary $\mx$ and $\my$,
the efficiency over the entire $(\mx,\my)$ plane is parameterized using the 
efficiencies calculated at the generated mass points.
As justified previously\cite{OPALSP183}, we restrict the searches to
$\mx+\my > M_{\rm Z}$
for the single-photon topology
and to $\mx$ values larger than about $M_{\rm Z}$/2 for the acoplanar-photons
topology.

\subsection{Single-Photon}
\label{sec:interpret_sp}

With the single-photon topology 
we measure the number of light
neutrino species and observe
a rise in the cross-section at low photon energies
which is consistent with the additional cross-section
expected from 
charged current contributions to 
$\ee \ra \nu_e \overline{\nu_e} \gamma (\gamma)$. 
We also give
results of searches 
for XY production,
superlight gravitino pair production, and
graviton-photon production in the context of models with extra space
dimensions.

\subsubsection{Neutrino Counting}
\label{sec:nnu}

   Single-photon events are expected within the Standard Model 
from the process $\ee \ra \nnggbra$. 
At the tree level, the cross-section 
for muon-type and tau-type neutrinos is attributable solely to
$s$-channel Z production 
with initial-state radiation. 
For electron-type neutrinos, the single photon 
cross-section arises from Feynman diagrams 
corresponding to $s$-channel Z production 
with initial-state radiation and $t$-channel W exchange with
radiation from the initial-state or the exchanged W.
Higher order electroweak processes such as WW boxes 
are expected to give a negligible contribution to 
the cross-section.
  
The results of the single-photon selection are used to measure the size 
of the $s$-channel Z production contributions and the W-related 
contributions (W amplitude squared plus the W-Z interference), which we 
parameterize in terms of $\rm{N_{\nu}}$ and $f_W$.
The measurement of the pure $s$-channel Z production contributions is
a direct measurement of the 
Z invisible width, 
which is related to the effective number of light neutrino generations, 
$\rm{N_{\nu}}$, defined as the ratio of the 
Z invisible width to the expected width in the Standard Model for one 
neutrino generation.  The W-related contributions are parameterized by 
a multiplicative scale factor, $f_W$, defined to be 1 for the Standard 
Model expectation.

In order to measure $\rm{N_{\nu}}$ and/or $f_W$,
we perform a binned fit to both the overall 
event rate and to the shape of the photon 
energy distribution 
by minimising the negative log-likelihood:
\begin{equation}
\label{eq:lik}
  -\log L = 
-\log {\cal P}(n_{exp}(E_{\gamma},{\mathrm{N_{\nu}}},f_W) \to n_{obs}(E_{\gamma}))
\end{equation} 
where $n_{exp}$ is given by
\begin{equation}
\label{eq:wtch}
{ n_{exp}(E_{\gamma}, {\mathrm {N_{\nu}}}, f_{W})} 
 = {f_{W} \; n_{exp}^{W} (E_{\gamma}) + {\mathrm{N_{\nu}}} \; n_{exp}^{Z} (E_{\gamma})}.  
\end{equation}
Here
$n_{obs}$
is the observed number of events, 
$n_{exp}$
is the expected number of events as a function of $\rm{N_{\nu}}$ and $f_W$
derived by reweighting 
fully simulated $\nnggbra$ events,
and ${\cal P}$ represents the Poisson probability 
for observing $n_{obs}$ events given an expectation of $n_{exp}$.  
$n_{exp}^{W}$ is the number of events 
for Standard Model W
contributions\footnote{Both the pure W $t$-channel, which is 
dominant,
and its interference with the $\rm{Z^{0}}$ $s$-channel are taken into 
account assuming Standard Model 
couplings of electrons and electron-type
neutrinos.}
and $n_{exp}^{Z}$ is the number of $s$-channel events 
expected per neutrino generation; both $n_{exp}^{W}$ and 
$n_{exp}^{Z}$ were evaluated using the NUNUGPV98 generator\footnote{
    Using the NUNUGPV98 feature which allows 
    $\rm{\nu_{e}\bar\nu_{e}\gamma (\gamma)}$ and 
    $\rm{\nu_{\mu}\bar\nu_{\mu}\gamma (\gamma)}$
    events to be generated separately
}.
 
The dominant systematic errors arise from the 
uncertainty on the selection efficiency 
and from the theoretical uncertainty 
on the expected number of events, with 
a minor contribution from the limited Monte Carlo statistics. 
The relative uncertainty on the selection efficiency is 2.1\% and
was discussed in Section~\ref{sec:results}.  
We assign a value of 2\% to the theoretical uncertainty
based on comparisons and estimated precisions of the NUNUGPV98, KORALZ
and \grcnng~\cite{bib:grcnunu} event generators. In addition,
an uncertainty related  to the modelling of the photon 
energy spectrum is assigned, again estimated by comparing 
the  NUNUGPV98 and KORALZ event generators.
Other sources of systematic error, such as the 
uncertainty on the centre-of-mass energy and the expected background, 
result in negligible contributions. 

In a first step, we check the consistency with the Standard Model
predictions by fitting for both 
$\rm{N_{\nu}}$ and $f_{W}$. The results are 
$f_{W}=1.12 \pm 0.13~\rm{(stat)} \pm 0.12~\rm{(sys)}$ and  
$N_{\nu}=2.63 \pm 0.15~\rm{(stat)} \pm 0.11~\rm{(sys)}$ 
with a correlation coefficient of $-41$\%.
Figure~\ref{f:sp_nnucont} shows the 70\%, 95\% and 99\% probability 
contours in the space of the two parameters, while 
Figure~\ref{f:sp_nnueg} 
shows the photon energy distribution for the 
events selected at 189 GeV,
compared to the expectation 
with $f_{W}=1.12$, $\mathrm{N_{\nu}}=2.63$.
The data are seen to be in fair agreement with the 
Standard Model 
prediction for $N_{\nu}=3$ and $f_W=1$.
In particular, the W contributions
are observed with a high degree of significance, even when 
allowing $N_{\nu}$ to be unconstrained.

In a second step, we assume the W contributions to be as 
predicted by the Standard Model and fit for $N_{\nu}$.
The result is: 
\begin{equation}
\label{eq:nnu}
N_{\nu} = 2.69 \pm 0.13~\rm{(stat)} \pm 0.11~\rm{(sys)}.
\end{equation}

Alternatively, one can assume $N_{\nu} = 3$
as indicated by the precise, but less direct, measurements
of the Z lineshape~\cite{bib:ls}, and fit for 
the relative size of
the W-contributions.
The result is $f_{W}=0.99 \pm 0.11~\rm{(stat)} \pm 0.12~\rm{(sys)}$,
establishing that the W-contributions are observed 
and are consistent with the expectations from the Standard Model.

%
%

\boldmath
\subsubsection{Search for $\eetoXY$, $\XtoYg$ ; General case: $\my\ge 0$}
\unboldmath
\label{sec:sp_results_allmy}

We search for evidence of new physics processes of type 
$\eetoXY$,~$\XtoYg$.
To select candidate new physics events,
kinematic-consistency and degraded-resolution cuts are applied
to events in the single-photon event sample.
The kinematic-consistency cuts require that the
energy of the most energetic photon be within the range 
kinematically consistent with mass values
$(\mx,\my)$
after accounting for energy resolution effects.
The degraded-resolution cuts reject events
in which the most energetic photon is in the angular
regions $0.78 < |\cos\theta| < 0.82$ or $|\cos \theta| > 0.95$; 
energy resolution is significantly degraded in these regions 
and the Monte Carlo simulation is less reliable. 
These cuts are approximately
94\% efficient in selecting signal events within the kinematic
acceptance of the single-photon selection, assuming uniform energy and
$\cos\theta$ signal distributions. 

Two methods are used to search for, and place upper limits on, contributions 
from $\eetoXY$,~$\XtoYg$.
The first is an event-counting method, based simply on the total
number of events selected as candidates.
The kinematic cuts, described above, are sufficiently loose that
the results are relatively insensitive to the shapes of 
signal distributions in photon energy or $\cos\theta$ that result 
from the specifics of a particular model.  Thus this method gives the 
more generally applicable cross-section limits for XY production.  
In many models, however, the distributions for the production and 
decay angles in the process $\eetoXY$,~$\XtoYg$ are approximately 
isotropic.  We therefore also perform
a likelihood-based analysis to search for new physics contributions
under the assumption of isotropically distributed production and
decay angles.  This results in a substantial increase in experimental
sensitivity, since energy and angular distributions can now
be used to differentiate signal from background.

In both methods, Standard Model background is assumed
to be from $\eetonnggbra$ only.  Background from other sources,
including the estimated cosmic ray and beam-related
background, is small and is neglected in the limit calculations.
Since uncertainties due to energy scale and resolution effects 
in the low $\mx - \my$ region 
lead to large relative uncertainties in the estimated efficiencies,
we restrict our search to the region in 
which $\mx - \my > 5$ GeV.

\paragraph{Event-Counting Method:\\}

Upper limits are calculated based on the total numbers of observed 
and expected background events using the method described in 
reference\cite{PDG96}.  This procedure is similar to the method used
in our previous publication\cite{OPALSP183}, with the exception that
only the kinematic-consistency and degraded-resolution
cuts are applied in order to improve the generality of the results.

  The total signal efficiency within the single-photon kinematic
acceptance is approximately 85\%, varying by less than 1\% over most
of the $(\mx,\my)$ plane.  However, the efficiency decreases
significantly when the mass difference $\mx - \my$ becomes
small, due to low photon energies resulting from the $\rm X \to Y \gamma$ 
decay.  
Contributions to the systematic error on the efficiency for selecting 
events from potential new physics sources are similar in nature and size
to those discussed in our previous publication. 
The total relative systematic error varies from 2.5 to 5.5\%,
depending on $\mx$ and $\my$.  
These systematic errors have been treated according to the method in 
reference\cite{systerr}; the effect on the upper limits is small.
Uncertainties in the $\nnggbra$ background estimate have a more 
significant effect on the upper limits, and are treated by means of a 
convolution within the limit calculations.  A relative 3\% uncertainty
has been assigned to the expected background contribution, based on
factors considered in the cross-section measurement and assigning 
a 2\% theoretical uncertainty.

There are 552 events selected after the degraded-resolution cuts are 
applied; the KORALZ expectation is $601 \pm 14$.  
The photon energy
distribution for these events is shown in Figure~\ref{f:sp_likeresult}.
The number of selected events after further application of
the kinematic-consistency cuts depends on the values for $\mx$ and $\my$;
ranges for observed and expected events 
are displayed as contour plots in 
Figure~\ref{f:sp_evsXY_data}.
The consistency between the number of observed events ($n_{obs}$) and 
the number of expected background events ($n_b$) is given in 
Figure~\ref{f:sp_consistency}.  Plotted is $P_{\rm fluct}$, the 
probability for observing at least $n_{obs}$ events given a background 
expectation of $n_b \pm \sigma_b$ events, defined by:
\begin{equation} 
  \label{eq:pfluct}
  P_{\rm fluct} =
  \sum_{n=n_{obs}}^{\infty} \left[ 
         \int_{-\infty}^{\infty} d\mu 
            \left( \frac{e^{(n_b-\mu)^2/2\sigma_b^2}}{\sqrt{2\pi\sigma_b^2}}
            \right)
            \cdot
           \left( e^{-\mu} \frac{\mu^n}{n!} \right) \right] .
\end{equation}
This figure is intended to highlight regions in the $(\mx,\my)$ plane
with an excess of events that could potentially be indicative
of new physics.  Such an excess would appear as a very low 
probability $(P_{\rm fluct} \ll 1\%$).  However, a deficit
of events is actually observed for most of the $(\mx,\my)$ plane.
This leads to high values for $P_{\rm fluct}$;
for example, for $n_{obs} \gg 1$, a value $P_{\rm fluct} = 0.97$
indicates a deficit consistent at approximately the 3\% level.
%
%
The calculated values of $P_{\rm fluct}$ range from 0.71\% to 99.94\%.

Values with rather low and high probability occur partly because
we consider several thousand points in the $(\mx,\my)$ plane; each
of these has a different set of kinematic-consistency cuts which
select different parts of the photon energy spectrum.  The net effect
is that almost every possible energy range
is selected for some $(\mx,\my)$ value.  For example, as seen in
Figure~\ref{f:sp_likeresult}, there is a 
deficit in the energy 
range from approximately 26 
to 38 GeV; this range corresponds roughly to $\mx=113$ GeV 
and $\my=74$ GeV.

The general structure of Figure~\ref{f:sp_consistency} can also be 
understood from the photon energy spectrum.
There is an overall deficit of approximately 49 events ($\sim 1.7 \sigma$).
As seen in Figure~\ref{f:sp_likeresult}, most of the deficit lies in 
the region of the radiative return peak, while there is an excess in the 
low-energy region.  For regions in the $(\mx,\my)$ plane for which
low-energy photons are not kinematically allowed, events populating
the low-energy region of the energy spectrum are rejected and the
significance of the remaining deficit is enhanced.
In particular, this occurs when $\mx - \my$ is relatively large,
and accounts for the large $P_{\rm fluct}$ values in the low-$\my$ region
of Figure~\ref{f:sp_consistency}.
Conversely, regions with a small mass difference $\mx - \my$ correspond 
to low photon energies, which in turn correspond to the region of the 
photon energy spectrum in which there is an excess.  This accounts 
for the small $P_{\rm fluct}$ values in the low $\mx - \my$ region of
Figure~\ref{f:sp_consistency}.

Cross-section times branching-ratio upper limits 
calculated using the event-counting method are shown
in Figure~\ref{f:sp_countlim}
and range from 43 to 409 fb.
Results for one $\mx,\my$ pair ($\mx = 125, \my= 63$ GeV) are shown 
as an addition to the expected $\nnggbra$ background
in Figure~\ref{f:sp_likeresult}.  The added signal contribution is that 
which would be expected from a cross-section equal to 
the 95\% CL upper limit.

\paragraph{Likelihood Method:\\}

The likelihood-based analysis is a straightforward extension of an 
extended maximum likelihood fit\cite{maxlike}.  Upper limits 
are calculated using information 
from the photon energy and angular distributions
as well as the total number of observed events.
The number of observed events, number of expected background events,
and signal efficiencies are the same as in the event-counting method.

The likelihood function is given by
\begin{equation}
  \label{eq:like}
  L(\sigma_s) = {\cal P}((n_b+n_s) \to n_{obs}) \cdot 
                \prod_i [f_b P_b(E_i, \theta_i) + f_s P_s(E_i, \theta_i)] 
\end{equation}
where the product is over all selected events in the data, and
\begin{itemize}
  \item $n_b$ = expected number of background events,
  \item $n_s = \epsilon_s \sigma_s {\cal L} = $
                expected number of signal events, with $\epsilon_s$
                the efficiency for observing signal events, 
                $\sigma_s$ the 
                signal cross-section times branching ratio, and ${\cal L}$ 
                the integrated luminosity,
  \item $n_{obs}$ = number of candidate events observed in the data,
  \item ${\cal P}((n_b+n_s) \to n_{obs}) = 
                e^{-(n_b+n_s)} (n_b+n_s)^{n_{obs}}/n_{obs}!$ = 
                the Poisson probability to observe $n_{obs}$ events
                given an expectation of $(n_b + n_s)$,
  \item $P_{b,s}(E_i, \theta_i)$ = probability density (normalized 
                to one) for a photon $i$ resulting from a background 
                ($b$) or signal ($s$) process to have an energy $E_i$ and 
                polar angle $\theta_i$, and
  \item $f_{b,s} = n_{b,s} / (n_b + n_s)$ = relative fraction of
                background or signal events.
\end{itemize}

The expected number of background events, $n_b$, is determined from 
a Monte Carlo sample of events generated by KORALZ.  
The signal efficiency, $\epsilon_s$, is calculated by integrating
an energy and angular distribution function over the region of 
kinematic acceptance (including kinematic-consistency and 
degraded-resolution cuts), and 
scaling the result to account for additional
efficiency losses due to other cuts in the single-photon selection.
This energy and angular 
distribution function serves as the the signal probability density 
function $P_s(E, \theta)$.  The background probability density 
function $P_b(E, \theta)$ 
is obtained from a parameterization of the distributions
formed by simulated $\nnggbra$ events generated by KORALZ 
and NUNUGPV98.  In the region $|\cos \theta| < 0.72$,
the photon energy and angular distributions are independent,
with the angular distribution given by 1/$\sin^2\theta$ and
the energy distribution determined with a parameterization.
In the region $|\cos \theta| > 0.72$,
energy resolution is dependent on $\cos\theta$.
In this region, energy distributions are
parameterized separately for slices of width 0.1 in $|\cos\theta|$; 
when properly normalized, these parameterizations together form the 
2-dimensional probability distribution.  

The value for $\sigma_s$ is not fixed; it is instead treated as
a free parameter in the likelihood function.
When properly normalized, 
the likelihood function can be thought of as a probability density function
for a hypothesized $\sigma_s$ consistent with the
observed data.  Restricting $\sigma_s$ to be non-negative, the 95\% CL upper 
limit $\sigma^{95}$ is therefore determined from the following equation:
%
%
\begin {equation}
  \label{eq:limit2}
  0.95 = \frac{\displaystyle \int_{0}^{\sigma^{95}} L(\sigma_s) d\sigma_s}
                {\displaystyle \int_{0}^{\infty} L(\sigma_s) d\sigma_s}.
\end{equation}


The validity of the likelihood-based method was tested with 
Monte Carlo simulations.  Two types of test were performed.  The
first test relies on the following definition of a 95\% CL limit:
if the true signal cross-section happened to 
be equal to the 95\% CL limit, then at least 95\% of a large number of 
identical experiments would result in data that are more signal-like 
(as defined by a likelihood comparison) than the experiment from which 
the 95\% CL likelihood was derived.  This type of test was performed
for various combinations of large and small numbers of expected 
signal and background events, using events generated randomly 
according to distributions derived from the expected XY signal and
$\nnggbra$ background.  The second type of test involved the
calculation of a large number of 95\% CL upper limits from 
samples generated with an identical signal cross-section.  In this 
case, at least 95\% of the calculated upper limits should be larger 
than the actual signal cross-section.  Both types of test 
yielded agreement with expectations to within statistical errors 
(typically less than 0.5\%).  
In addition, tests with fully simulated
Monte Carlo events were used to test for any biases due to the
signal and background parameterizations.  No significant biases were
found.
 
Sources of systematic errors are the same as for the event-counting
method, and are treated in the same manner.  However, when calculating 
limits using the likelihood method,
an additional 4\% relative uncertainty in experimental sensitivity 
has been added to 
account for effects due to uncertainties in the background and signal 
parameterizations.  The 4\% estimate is the result of tests using
different parameterizations; the uncertainty is again treated according 
to the method given in reference\cite{systerr}.

The 95\% CL upper limits 
resulting from the likelihood method are shown 
as a function of $M_{\rm X}$ and $\my$ 
in Figure~\ref{f:sp_likelim}.  The values range from 23 fb to 371 fb.


\boldmath
\subsubsection{Search for $\eetoXY$, $\XtoYg$ ; Special case: $\myzero$}
\unboldmath
\label{sec:sp_results_my0}

The case $\myzero$ is applicable to excited
neutrino models and to some supersymmetric models mentioned earlier.
The results presented above include this case and no separate
analysis is performed, although the results are highlighted here.
The upper limits on $\sigbrXY$ for $\myzero$ as a function of $\mx$
range from 23 to 107~fb using the likelihood method, and from 
43 to 170~fb using the event-counting method.
The limits are shown in Figure~\ref{f:sp_XYlim_massless} for
both methods.  Also shown is the average limit expected in the
absence of signal for the event-counting method. 
The actual limit
is significantly lower than the expected limit (see discussion in  
Section~\ref{sec:sp_results_allmy}).
Table~5
gives more details of the search results.
%

\boldmath
\subsubsection{Gravitino Pair Production}
\unboldmath
\label{sec:sp_results_GGg}

A supersymmetric model has been proposed in 
which the gravitino is very light\cite{Zwirner}.  
This model predicts a new source of single-photon events 
from the process $\epem \ra \Gravitino \Gravitino \gamma$. 
We use the single-photon topology to place constraints on the 
cross-section and therefore the gravitino mass in this model.
The differential cross-section is given by
\begin{equation}
  \frac{d^2\sigma}{dx_{\gamma}\,d\cos\theta} = 
  \left( \frac{\alpha {G_N}^2}{45} \right) 
  \frac{s^3}{{m_{\gv}}^4} f_{\gv\gv\gamma}(x_{\gamma},\cos\theta)
  \label{eq:GGg1}
\end{equation}
where $\alpha$ is the fine structure constant, $G_N$ is the 
gravitational constant, $m_{\gv}$ is the gravitino mass, 
$x_{\gamma}$~is the photon scaled energy ($E_\gamma/E_{\rm beam}$) 
and $\theta$ is the polar angle, with
\begin{equation}
  f_{\gv\gv\gamma}(x,\cos\theta) = 
        2(1-x)^2 \left[ \frac{(1-x)(2-2x+x^2)}{x\sin^2\theta} +
                \frac{x(-6+6x+x^2)}{16} - \frac{x^3\sin^2\theta}{32} \right],
  \label{eq:GGg}
\end{equation}
leading to a soft photon energy spectrum.

Two methods were used to determine 
the efficiency for observing events from $\epem \to \gv \gv \gamma$.
The first is by reweighting simulated $\nnggbra$ events generated by KORALZ.
Each simulated event is weighted by 
$f_{\gv\gv\gamma}(x, \cos\theta) / f_{KZ}(x, \cos\theta)$,
where $f_{KZ}$ is the 2-dimensional energy and angular distribution for the 
most energetic photon in events generated by KORALZ.
%
We use a parameterization $f_{KZ}$ which is valid up to photon 
energies of about 
60 GeV, and thus we establish a kinematic acceptance region for 
$\gv \gv \gamma$ production: $x_T > 0.05$, $15^\circ < \theta < 165^\circ$,
and $E_\gamma < 60$ GeV.  We note that the cross-section for 
$\gv \gv \gamma$ production with photon energies above 60 GeV is
negligible.

In addition, a Monte Carlo generator was written to generate
events according to the distribution given in Equation~\ref{eq:GGg}. 
Initial state radiation 
was treated in a manner identical to that used by the EXOTIC\cite{EXOTIC} 
generator.  Efficiencies were calculated and found to be identical to 
those using the reweighting method to within statistical errors.

To achieve maximum sensitivity\footnote{
        The optimization condition
        chosen was that the expected upper limit on the
        cross-section for contributions from the
        $\epem \to \gv \gv \gamma$ signal process be minimised,
        where the expected upper limit is defined as the 
        average limit one would expect to set in the
        absence of signal. },
we place a maximum energy requirement on the observed photon 
energy: $E_\gamma < 30 $ GeV.  With this requirement, we observe
195 candidates with an expected $\nu \bar{\nu} \gamma$ 
background of $179.6 \pm 5.4$.  The efficiency within the 
above kinematic acceptance region is estimated by the Monte
Carlo to be 82.8\%, and was assigned a relative 5\% 
uncertainty.  Using the event-counting method described earlier,
we place a 95\% CL cross-section upper limit of 293 fb, giving a 
lower limit\footnote{Evaluated with $\alpha =
\frac{1}{128}$}
 on the gravitino 
mass of 8.7 $\mu$eV.

\boldmath
\subsubsection{Graviton-Photon: Search for Extra Space Dimensions}
\unboldmath
\label{sec:sp_results_Gg}

There has been recent interest in string theory models which
postulate the existence of additional compactified space dimensions; 
these models allow gravitons to propagate freely in the higher-dimensional 
space while restricting Standard Model particles to a 3+1 dimensional 
hypersurface\cite{ADD}.  
%
%
The fundamental mass scale in this class of theories, $M_D$, governs 
the rate
of graviton production;
it is possible that direct graviton-photon production could occur at 
significant rates at LEP2 energies with
an experimental signature of a single photon with missing 
energy\cite{Peskin,GRW}.  The single-photon search topology can therefore 
be used to place constraints on the fundamental mass scale $M_D$
(or, equivalently, on the size of the extra dimensions) by placing
limits on the graviton-photon cross-section.  

The differential cross-section is given by
\begin{equation}
  \frac{d^2\sigma}{dx_{\gamma}\,d\cos\theta} = 
  \frac{\alpha S_{\delta-1}} {64 {M_D}^2 }
  \left(\frac{\sqrt{s}}{M_D} \right)^{\delta} 
   f_{G \gamma}(x_{\gamma},\cos\theta)
  \label{eq:Gg1}
\end{equation}
where $\delta$ is the number of extra dimensions and
$S_{\delta-1}$ is the surface area of a $\delta$-dimensional
sphere of unit radius, 
with

\begin{equation}
   f_{G\gamma}(x,\cos\theta) = 
        \frac{2 (1-x)^{\frac{\delta}{2} - 1}}{x(1-\cos^2\theta)}
        \left[(2-x)^2 (1-x+x^2) - 3x^2 \cos^2\theta (1-x) - 
                x^4 \cos^4\theta \right].
  \label{eq:ggamma}
\end{equation}



We use the event-counting method 
to place limits on 
graviton-photon production in the cases $2 \le \delta \le 7$.
We use
the same kinematic acceptance region as described
in Section~\ref{sec:sp_results_GGg}: $x_T > 0.05$, 
$E_\gamma < 60$~GeV, and $15^\circ < \theta < 165^\circ$.
The expected photon energy spectrum is soft, so in order
to improve sensitivity, we require observed photon energies
to be less than 34 GeV for all values of $\delta$.

The efficiency for observing events from $\epem \to G \gamma$
is determined by reweighting simulated $\nnggbra$ events generated
by KORALZ in a manner equivalent to that described in 
Section~\ref{sec:sp_results_GGg}:
simulated events are weighted by 
$f_{G\gamma}(x, \cos\theta) / f_{KZ}(x, \cos\theta)$,
where $f_{KZ}$ is the same KORALZ parameterization used earlier.
The distributions $f_{KZ}$ and $f_{G\gamma}$ have roughly similar shapes, 
and the calculated 
efficiency is insensitive to small differences in the KORALZ
parameterization.

We observe 208 candidates with an expected $\nu \bar{\nu} \gamma$ 
contribution of $196.0 \pm 5.9$ events.
Assuming a 5\% relative uncertainty in the signal efficiency, 
upper limits on the cross-section and corresponding lower limits on
$M_D$ for $2 \le \delta \le 7$ have been 
calculated\footnote{Evaluated also with $\alpha =
\frac{1}{128}$}.  
The 
results are given in 
Table~6
based on the convention\footnote{The convention of reference\cite{Peskin}
differs.} of Equation 2
in reference\cite{GRW}.

\subsection{Acoplanar-Photons}

\boldmath
\subsubsection{Search for $\eetoXX$, $\XtoYg$ ; General case: $\my\ge 0$}
\unboldmath
\label{sec:g2_results_allmy}

The searches for $\eetoXX$, $\XtoYg$, both for the general case discussed
here and the special case of $\myzero$ discussed in 5.2.2, use
the methods described in our previous publication\cite{OPALSP183}.
Selected events are classified as consistent with a given value of $\mx$
and $\my$ if the energy of each of the photons falls within the region
kinematically accessible to photons from the process $\eetoXX$,
$\XtoYg$, including resolution effects. 
The selection efficiencies at each generated grid point for the 
$\eetoXX$, $\XtoYg$ Monte Carlo events 
are shown in Table~\ref{tab:g2_eff_189}. These values include the
efficiency of the kinematic consistency requirement which is higher than
95\% at each generated point in the $(\mx,\my)$ plane for which 
$\mx-\my > 5$ GeV.
Events from $\eetonngggbra$ are typically characterized by a high-energy 
photon from the radiative return to the $\rm Z^0$ and a second lower energy
photon. The kinematic consistency requirement is such that the two
photons must have energies within the same (kinematically accessible)
region. As $\mx$ and $\my$ increase, the allowed range of energy for the
photons narrows, and fewer $\nngggbra$ events will be accepted. For the
24 selected events, the distribution of the number of events consistent with 
a given mass point ($\mx$,$\my$) is consistent
with the expectation from $\eetonngggbra$ Monte Carlo, over the full
($\mx$,$\my$) plane.

In our previous publication, because of concerns about the modelling of
the Standard Model process $\eetonngggbra$, all limits derived from the
acoplanar photons analysis were obtained without accounting for the 
expected background. The theoretical situation is now greatly improved,
with two event generators\cite{NUNUGPV98,KORALZ} agreeing to better
than 1\% for the total cross-section for this process within the kinematic
acceptance of this analysis. For that reason, in this paper, all limits
derived from this selection have been calculated taking the (KORALZ) background
estimate into account. Figure~\ref{f:g2_mxmy_189} shows the 95\% CL exclusion regions for
$\sigbrXX$. The limits vary from 33 fb to 103 fb for $\mx >45$ GeV
and $\mx - \my >5 $ GeV. In the region 2.5 GeV $\le \mx -\my < 5.0$ GeV, the 
efficiency falls off rapidly (see Table~\ref{tab:g2_eff_189}). As this rapid
fall increases the associated uncertainty in the efficiency, no limits have
been set in this region.

%

Systematic errors are due primarily to limited Monte Carlo statistics at
the generated ($\mx,\my$) points and the uncertainty on the efficiency
parameterization across the ($\mx,\my$) plane. The combined relative
uncertainty on the efficiency varies from about (3-6)\% across the plane
(for $\mx - \my > 5$ GeV). All systematic uncertainties are accounted
for in the manner advocated in reference\cite{systerr}. This also
applies to the limits for the $\myzero$ case, presented in the next
section.

\boldmath
\subsubsection{Search for $\eetoXX$, $\XtoYg$ ; Special case: $\myzero$}
\unboldmath
\label{sec:g2_results_my0}

For the special case of $\myzero$ the kinematic consistency requirements
differ from those used for the general case. One can
calculate\cite{gravitinos2} the maximum mass, $\mxmax$, which is
consistent with the measured three-momenta of the two photons, assuming
a massless $\PY$.
A cut on $\mxmax$  provides further suppression of the $\nngggbra$ 
background while retaining high efficiency for the  signal hypothesis. 
This is discussed in more detail in reference\cite{OPALSP172}.
We require that the maximum kinematically allowed mass be 
greater than $\mx-5$ GeV,
which retains $(95.5^{+2.0}_{-1.0})$\% relative efficiency 
for signal at all values 
of $\mx$ while suppressing much of the remaining $\nngggbra$ background. 
In our previous publication we also applied a recoil-mass cut at 80 GeV.
In the case where background is not accounted for in the limit calculation,
such a cut improves the expected sensitivity of the analysis. With background
subtraction this is no longer the case. Therefore that cut has 
been removed.

Figure~\ref{f:g2_mxmax_189} shows the expected $\mxmax$ distribution 
for signal Monte Carlo events with $\mx = 90$ GeV and for $\eetonngggbra$ Monte
Carlo events.  Also shown is the distribution of the selected data events.
For the $\myzero$ case, the efficiencies calculated from Monte Carlo
events are shown in Table~\ref{tab:g2_eff_my0_189} 
after application of the event selection criteria and then after the cut on 
$\mxmax$. Also shown in 
Table~\ref{tab:g2_eff_my0_189} are the number of selected events consistent
with each value of $\mx$ as well as the expected number of events from
$\eetonngggbra$. The number of selected events consistent with a given 
value of $\mx$ varies from 14, for $\mx\ge$ 45 GeV, to 3 events at the kinematic 
limit. The expected number of events decreases from $15.8 \pm 0.7$ at 
$\mx\ge 45$ GeV to $1.34 \pm 0.07$ consistent with $\mx \ge 94$ GeV.

Based on the efficiencies and the number of selected events, we calculate a 
95\% CL upper limit on $\sigbrXX$ for $\myzero$ 
as a function of $\mx$. 
This is shown as the solid line in Figure~\ref{f:g2_limit_my0_189}.
The limit is between 50 and 80 fb for $\mx$ values from 45 GeV up 
to the kinematic limit. Also shown as a dashed line is the expected limit, 
defined as the average limit one would expect to set in the absence
of signal.
The limits can be used to set model-dependent limits on the mass of the 
lightest neutralino in supersymmetric models in which the 
NLSP is the lightest neutralino and the
LSP is a light gravitino ($\PX=\lsp, \PY=\Gravitino$).
Shown in Figure~\ref{f:g2_limit_my0_189}
as a dotted line is the (Born-level) cross-section prediction from
a specific light gravitino LSP model\cite{chang} in which
the  neutralino composition is purely bino, 
with $m_{\tilde{e}_R} = 1.35 m_{\lsp}$ and $m_{\tilde{e}_L} = 2.7 m_{\lsp}$.
Within the framework of this model, $\lsp$ masses between 45 and 88.3 GeV
are excluded at 95\% CL. 

As described in section 2, the efficiencies over the full angular range have
been calculated using  isotropic angular distributions for production and 
decay of $\PX$. The validity of this model  has been examined
based on the angular distributions calculated for photino pair 
production in reference\cite{ELLHAG}. For models proposed 
in reference\cite{gravitinos}, the 
production angular distributions are more central and so this procedure is 
conservative. For a $1 + \cos^2{\theta}$ 
production angular distribution expected for t-channel exchange of a 
very heavy particle according to reference\cite{ELLHAG}, the 
relative efficiency reduction would be less than 2\% at all
points in the ($\mx,\my$) plane.

\boldmath
\subsubsection{Search for $\eetoXX$, $\XtoYg$ ; Special case: $\myzero$ and
macroscopic decay length}
\unboldmath
\label{sec:g2_results_my0_lifetime}
As an extension to the  special case of $\myzero$, we consider 
the sensitivity 
of the acoplanar-photons search to $\eetoXX$, $\XtoYg$ when $\PX$ has a 
macroscopic decay length. 
This extension 
evaluates the selection efficiency using 
signal Monte Carlo samples with various lifetime values $\tau_X$.  
The $\XtoYg$ decay is 
treated by a modified
OPAL detector simulation package designed to 
handle delayed decays (as described in section \ref{sec:detector}). 
The range of lifetimes considered extends 
from the near-zero lifetime 
value of 
$\tau_X=10^{-15}$~s
to 
$\tau_X=10^{-7}$~s
at which point c$\tau$
is 30 m. 
Such long lifetimes lead to decays outside the 
sensitive volume of the OPAL detector (determined by the outer radius of the 
ECAL), which implies a natural cutoff in sensitivity.

As with the zero lifetime case, the ($\mxmax > \mx -5$ GeV) requirement 
is applied.
The relative efficiency of this cut
decreases with increasing lifetime, but exceeds 90\% 
for all $(\mx,\tau_X)$ combinations with high (greater than 50\%)
selection efficiencies.
For $(\mx,\tau_X)$ combinations with lower selection efficiencies, the
relative efficiency falls more rapidly with increasing $\tau_X$, but 
does not drop below 60\%
for the  $(\mx,\tau_X)$ combinations considered.
The loss in relative efficiency of the $\mxmax$
cut is expected due to the definition of $\mxmax$, but the cut is 
maintained to allow comparison 
between the prompt decay and macroscopic decay length 
selection 
efficiencies. The resulting selection efficiencies for macroscopic
decay lengths
are given in  
Table \ref{tab:g2_eff_my0_189_np}.
Also listed in Table \ref{tab:g2_eff_my0_189_np}  are the number of events 
observed in the data and the expected $\eetonngggbra$ background.
The efficiencies indicate that 
sensitivity to 
macroscopic decay lengths is maintained 
up to 
lifetimes of
$10^{-9}$~s
but falls rapidly for 
longer lifetimes. This is as expected, 
since for 
$\tau_X=10^{-9}$~s
the decay 
length ranges from 30 to 3 cm for the values of $\mx$ 
considered, so that the $\XtoYg$ decays typically occur well 
within the confines of the detector. 
For lifetimes greater than $10^{-9}$~s, 
the
decrease in efficiency is two-fold.  The primary loss is
due to the increased decay length, implying fewer $\XtoYg$ decays within 
the sensitive detector volume. Selection 
efficiency is also lost with 
increasing $\tau_X$ since larger lifetimes result in a 
delayed arrival time of the photon at the ECAL, 
which, if sufficiently late, causes 
the event to be vetoed  due to the  timing cuts imposed by 
the analysis. The effect of the event kinematics is 
also seen in the increase in selection efficiency 
for a given lifetime as $\mx$ approaches its 
threshold value, corresponding to a drop in $\beta$ 
from 0.84 at $\mx$=50 GeV  to 0.10 at $\mx$=94 GeV.  

Contributions to the systematic error in the selection
efficiency are (in order of significance)
the limited Monte Carlo statistics, the efficiency parameterisation 
in the $(\mx,\tau_X)$ plane
and 
the modelling of the timing cuts.
The total contributions are given in Table~\ref{tab:g2_eff_my0_189_np}.

Given the efficiencies and the number of selected events, the 95\%
CL upper limit on $\sigbrXX$ for $\myzero$ 
as a function of $\mx$ and $\tau_X$ is calculated, and is shown in figure 
\ref{f:g2_limit_mx_tau_pr189}. The background-subtracted
limits from the prompt decay analysis are used to define the profile as a 
function of $\mx$ for $\tau_X=10^{-15}$~s. The interpolation of the 
limit to larger lifetimes is done by factoring the selection efficiencies of 
Table~\ref{tab:g2_eff_my0_189_np} into the 95\% 
confidence level limits on $\sigbrXX$ for $\myzero$.  
As in Section {\ref{sec:g2_results_my0} the Born-level 
cross-section of the neutralino 
NLSP Gravitino LSP model of  \cite{chang} is used to give 
an exclusion region in 
the (neutralino mass, lifetime) plane. This exclusion region 
is superimposed on 
Figure \ref{f:g2_limit_mx_tau_pr189} 
and gives the excluded domain in the ($\mx$,$\tau_X$) plane
within the same specific light gravitino LSP model \cite{chang} 
discussed above.

\section{Conclusions}
We have searched for photonic events with large missing energy in two 
topologies in data taken with the OPAL detector at LEP, 
at a centre-of-mass energy of 189 GeV.

%
%
In the single-photon selection, which requires at least one photon with
$x_{T} > 0.05 $ in the region $15^{\circ}<\theta<165^{\circ}$
($\acosthe < 0.966)$, 643 events 
are observed in the data.
The background-subtracted cross-section measurement of
4.35 $\pm$ 0.17 (stat) $\pm$ 0.09 (sys)~pb 
is consistent with the KORALZ prediction of 4.66 pb
from the Standard Model $\nnggbra$ process.
Interpreting the results as a measurement of
the effective number of light 
neutrino species, we measure
$\rm N_{\nu} = 2.69 \pm 0.13 (stat) \pm 0.11 (sys)$.
We also observe significant W contributions to the cross-section 
with a rate consistent with the Standard Model expectation.

We calculate upper limits on the 
cross-section times branching ratio for the process $\eetoXY$, $\XtoYg$
using two methods:
an event-counting method, which is insensitive
to energy or angular distribution shapes and is therefore relatively
model-independent, and a likelihood method, which assumes isotropic
production and decay angular distributions and has greater 
sensitivity.
In the region of interest in the $(\mx,\my)$ plane,
the limits vary from 43 to 409 fb using the event-counting method
and from 23 to 371 fb using the likelihood method.
These limits include 
the special case of $\myzero$, where 
the limit varies between 43 and 170 fb using the event-counting method
and from 23 to 107 fb using the likelihood method.
We note that some of these limits are much more stringent than
would be expected on average in the absence of signal contributions.

We
set a 95\% CL cross-section upper limit on $\ee \ra \gv \gv \gamma$
production of 293 fb
implying a lower limit on the gravitino 
mass of 8.7 $\mu$eV in the superlight gravitino model of 
reference\cite{Zwirner}.
In the context of string theory models with extra space dimensions,
upper limits on the cross-section for graviton-photon
production of between 309 and 271 fb
at 95\% CL are set, giving lower limits on the fundamental mass scale 
varying between 1086 and 470 GeV for between 2 and 7 additional dimensions
respectively. 

The acoplanar-photons selection requires 
at least two photons with scaled energy
$x_{\gamma}>0.05$ within the polar angle 
region $15^{\circ}<{\theta}<165^{\circ}$
or at least two photons with energy $E_{\gamma}>1.75$ GeV with 
one satisfying
$\acosthe < 0.8$ and the other 
satisfying $15^{\circ}<{\theta}<165^{\circ}$.
In each case, the requirement 
$p_T^{\gamma\gamma}/E_{\rm beam} > 0.05$
is also applied. There are 24 events selected.
The KORALZ prediction for the contribution from 
$\eetonngggbra$ is $26.9 \pm 1.2$ events; contribution from 
other sources is 0.11 events.
The number of events observed in the data
and their kinematic distributions are consistent
with Standard Model expectations. 
We derive 95\% CL upper limits on 
$\sigbrXX$ ranging from 33 to 103 fb for the general case of massive 
$\PX$ and $\PY$. 
For the special case of $\myzero$, 
the 95\% CL upper limits on $\sigbrXX$ range from 50 to 80 fb.
For the case of macroscopic decay lengths, these values 
range from 50 fb at short lifetimes to 8.8 pb for lifetimes 
as long as $10^{-7}$ s.

The results of the 
acoplanar-photons search are used to place
model-dependent lower limits on the $\lsp$ mass 
in a specific light gravitino LSP model\cite{chang}.
Masses between 45 and 88.3 GeV are excluded at 95\% CL for 
promptly decaying neutralinos, while for a $\lsp$ lifetime of
$10^{-8}$ s ($c\tau = 3$m), masses between 45 and 81 GeV are excluded.

\newpage
\noindent {\Large\bf Appendix}

\appendix
\renewcommand{\thesection}{\Alph{section}}

\section{Improvements to Event Selections}
Detailed descriptions of the 
event selection criteria are given in a previous 
publication\cite{OPALSP183}.  
This appendix describes changes and updates to the selections used 
in the present publication.

\subsection{Single-Photon Selection}
There have been several improvements to the single-photon selection:

\begin{itemize}

\item Non-conversion candidates in the endcap region ($|\cos\theta|>0.82$) 
are no longer required to have a good in-time associated TE hit, 
although candidates with associated out-of-time TE hits are 
still rejected.
This change reduces sensitivity to the modelling of the material
between the interaction point and the TE scintillators, and
results in a relative increase in efficiency of about 19\%
in the endcap region with an associated increase in cosmic and
beam-related background of about 1.3 events.  
The special
background vetos continue to be applied to all non-conversion candidates
in the endcap regardless of the presence of TE timing.  

\item The cluster extent cut has been modified for events in the
endcap region.  This cut had been fixed at lower centre-of-mass
energies and was no longer fully efficient 
at $\roots=189$ GeV,
particularly for beam energy photons.
Events with 
the primary photon candidate in the region 
$\acosthe > 0.82$ are now
rejected if $\Delta \phi \cdot \sin^{1.46}\theta/\ln E_\gamma > 0.05$
radians\footnote{
	The $\sin\theta$ dependence was parameterized for beam energy
	photons using samples of real and simulated 
	$\ee \to \gamma \gamma$ events.},
where $\Delta \phi$ is the $\phi$ extent of the cluster and 
$E_\gamma$ is the cluster energy in GeV.
The result is a 
relative increase in efficiency of about 13\% in the endcap region with 
no significant change in 
expected cosmic and beam-related background.

\item 
Small changes have been made to the timing requirements.
The timing cuts for the photon candidates
with TOF associated hits ($\acosthe < 0.82$) have been relaxed
for certain periods of data-taking, 
resulting in an efficiency increase of 2.3\% for events with a photon 
in $\acosthe < 0.82$
with a TOF associated hit.
The expected 
cosmic ray background increases by about 10\%.

In addition, the majority of events rejected by poorly measured
TOF times are recovered.
In cases where the ECAL cluster and the TOF z-measurement from time
difference
differ by more than 40 cm,
modified arrival times
based on the ECAL cluster position and the time measurement
at each end of the scintillator are constructed. 
Events are retained if either modified arrival time is 
within 5 ns
of the expected time for a photon originating from the interaction
point.  As a result, the probability for a signal event to be rejected due to
bad TOF timing falls from 0.6\% to less than 0.1\% with no measurable
change in the cosmic and beam-related background.

\item The efficiency of the photon conversion 
consistency criteria has been improved 
substantially. 
In particular, all reconstructed tracks in the 
event (usually two for signal events)
are now used to test association of a charged track to an ECAL
cluster, and the association test also considers 
the disfavoured solution of the jet-chamber left-right ambiguity
to account for cases where the incorrect solution has been
chosen.

\item 
Improved redundancy in the rejection of beam-related backgrounds
for conversion candidates with 
$x_T < 0.1$ and $|\cos{\theta}| > 0.90$ is introduced, 
based on the measured z-difference of the track's point of closest approach
to the interaction point.


\item 
An additional beam-halo veto was 
implemented; events are rejected if ECAL 
clusters exceeding 100 MeV in deposited energy are found within 20 cm
in the $r-\phi$ plane but on opposite ends of the
ECAL endcap.  
This veto provides better rejection of beam-halo background 
and has negligible effect on signal efficiencies.  

\item 
A minor addition was made to the special background vetoes used
for non-conversion candidates.
Events are now explicitly rejected if there is a large amount of 
hadronic energy in an HCAL cluster associated with the ECAL cluster
of the primary photon candidate.  
This provides better redundancy with existing cuts in terms of background
rejection, with negligible inefficiency even for beam energy photons.

\end{itemize}

\subsection{Acoplanar-Photons Selection}

The acoplanar-photons selection is almost identical to that described
in our previous publication.  The only difference involves the timing 
requirements for candidates with associated TOF hits.  These were
relaxed for certain periods of the data-taking in the same manner as
described above for the single-photon selection.

\section*{Acknowledgements}
The authors wish to thank G. Giudice, M. Peskin and F. Zwirner
for helpful clarifications.

We particularly wish to thank the SL Division for the efficient operation
of the LEP accelerator at all energies
 and for their continuing close cooperation with
our experimental group.  We thank our colleagues from CEA, DAPNIA/SPP,
CE-Saclay for their efforts over the years on the time-of-flight and trigger
systems which we continue to use.  In addition to the support staff at our own
institutions we are pleased to acknowledge the  \\
Department of Energy, USA, \\
National Science Foundation, USA, \\
Particle Physics and Astronomy Research Council, UK, \\
Natural Sciences and Engineering Research Council, Canada, \\
Israel Science Foundation, administered by the Israel
Academy of Science and Humanities, \\
Minerva Gesellschaft, \\
Benoziyo Center for High Energy Physics,\\
Japanese Ministry of Education, Science and Culture (the
Monbusho) and a grant under the Monbusho International
Science Research Program,\\
Japanese Society for the Promotion of Science (JSPS),\\
German Israeli Bi-national Science Foundation (GIF), \\
Bundesministerium f\"ur Bildung, Wissenschaft,
Forschung und Technologie, Germany, \\
National Research Council of Canada, \\
Research Corporation, USA,\\
Hungarian Foundation for Scientific Research, OTKA T-029328,
T023793 and OTKA F-023259.


\newpage
%
%
\begin{table}[b]
\centering
\begin{tabular}{|c|c|c|} \hline
Quantity & \multicolumn{2}{c|}{Topology} \\ \hline
 ~       &  Single-photon & Acoplanar photons \\ \hline \hline
$\rm N_{obs}$        & 643  & 24 \\ \hline
${\rm N_{expected}}$ & $679 \pm 5 \pm 14$ &  $26.9 \pm 1.2$ \\ \hline
$\rm N_{bkg}$        & $9.2 \pm 1.6$ & $0.11 \pm 0.04$ \\ \hline
Efficiency (\%) &  $82.1 \pm 1.7$  & $66.4 \pm 2.9$ \\ \hline
$\rm {\sigma}_{meas}$(pb) & 
        $4.35 \pm 0.17 \pm 0.09$ & $0.204 \pm 0.043$ \\ \hline
$\rm {\sigma}_{KORALZ}$(pb) & 
        4.66 $\pm$ 0.03 (stat) & $0.228 \pm 0.002$ (stat) \\ \hline
\end{tabular}
\caption{Summary of results for the single and acoplanar photons
selections.
Shown are the number of events observed in the data; 
the number of expected $\eetonnggbra$ (single-photon) or $\eetonngggbra$ 
(acoplanar-photons) events based on the KORALZ event generator; 
the number of events expected from backgrounds;
the selection efficiency within the kinematic acceptance of the selections
(defined in section 3);
the background-subtracted, measured cross-section
within the kinematic acceptance; and the
expected cross-section based on the
KORALZ generator.
Errors when not marked are statistical and systematic added
in quadrature.
When two errors are shown, the first is statistical and the second
is systematic.
}
\label{tab:sp_results}
\end{table}
%

\begin{table}[b]
\centering
\begin{tabular}{|l||c|} \hline
Background process &
 Number of events   \\ \hline \hline
$\ee \ra \ee\gamma$ & 0.3 $\pm$ 0.2  \\ \hline 
$\ee \ra \ell^{+} \ell^{-} \nu\bar{\nu} (\gamma)$ & 1.80 $\pm$ 0.14  \\ \hline 
$\rm \ee \ra \ee \ell^+ \ell^- $ & 1.2 $\pm$ 0.4 \\ \hline
$\rm \ee \ra \nu \bar{\nu} q\bar{q}$ & 0.71 $\pm$ 0.09 \\ \hline
$\eetomumu(\gamma)$ & 0.37 $\pm$ 0.08  \\ \hline 
$\eetotautau(\gamma)$ & 0.17 $\pm$ 0.05  \\ \hline 
$\eetogg(\gamma)$ &  0.0 ($<0.05$)  \\ \hline \hline
Total physics background & 4.6 $\pm$ 0.5  \\ \hline 
Cosmic background  &  2.6 $\pm$ 0.6 \\ \hline
Beam related background & 2.0 $\pm$ 1.4 \\ \hline \hline
Total background & 9.2 $\pm$ 1.6  \\ \hline 
\end{tabular}
\caption{Numbers of events expected from various
background processes contributing to the 
single-photon event sample.
The Standard Model background contributions
are given by process.
Also shown are the expected other backgrounds 
from cosmic rays and beam related sources. 
The errors shown are statistical.
}
\label{tab:sp_background}
\end{table}

\begin{table}[b]
\begin{center}
\begin{tabular}{|l|c|}\hline
Systematic  & Relative Error(\%) \\ \hline
Selection Efficiency & 1.5 \\
Occupancy probability & 1.0 \\
Early conversion & 0.7 \\
Tracking systematics & 0.5 \\
Event generator physics modelling & 0.5 \\
$x_T$ scale and resolution & 0.4 \\
Integrated luminosity & 0.2 \\
Angular acceptance   & 0.2 \\ 
MC statistics & 0.2 \\ \hline
Total & 2.1 \\ \hline
\end{tabular}
\caption{\small Summary of the experimental systematic errors related to
the efficiency and normalisation of the single photon
cross-section measurement.}
\end{center}
\label{tab:sp_systematics}
\end{table}

\begin{table}
\centering
\begin{tabular}{|c||c|c|} \hline
$\roots$(GeV) & $\mathrm {\sigma}_{meas}^{\nngggbra}$(pb) & 
$\mathrm {\sigma}_{KORALZ}^{\nngggbra}$(pb) \\ \hline \hline
130 & $1.49\pm{0.68}$  & $0.626\pm{0.010}$\\ \hline 
136 & $1.23\pm{0.56}$  & $0.526\pm{0.008}$ \\ \hline 
161 & $0.16\pm{0.16}$  & $0.330\pm{0.018}$ \\ \hline
172 & $0.32\pm{0.23}$  & $0.303\pm{0.017}$ \\ \hline
183 & $0.27\pm{0.09}$ & $0.247\pm{0.002}$ \\ \hline 
189 & $0.20\pm{0.04}$ & $0.228\pm{0.002}$ \\ \hline
\end{tabular}
\caption{The measured cross-section for the process $\eetonngggbra$,
within the kinematic acceptance defined in section 3, for different
centre-of-mass energies. For $\roots =$ 130 and 136 GeV the measurements
are the weighted average of the results obtained from the 1997 data and
the results obtained from the 1995 data. 
Results for $\roots \le$ 183 GeV are taken from our previous
publications\protect\cite{OPALSP183,OPALSP172}.
The final column shows the cross-section predictions from KORALZ.
The quoted errors are statistical.
}
\label{tab:g2_xsec_189}
\end{table}

\begin{table}[b]
\begin{center}
  \begin{tabular}{|c|c|c|c|c|c|c|c|} \hline

   $\mx$ & $E_{min}$ & $E_{max}$ & $\rm N_{obs}$ & $\rm N_{expected}$
        & $P_{\rm fluct}$ & $N^{95}$ & $\langle N^{95} \rangle$ \\ \hline
   \hline
    100 & 25.1 &  97.6 & 390 & $449 \pm 13$ & 0.9921 & 23.4 & 52.6 \\ \hline
    110 & 30.5 &  97.7 & 371 & $428 \pm 13$ & 0.9917 & 22.9 & 51.0 \\ \hline
    120 & 36.3 &  97.9 & 358 & $407 \pm 12$ & 0.9832 & 24.1 & 49.5 \\ \hline
    130 & 42.6 &  98.1 & 342 & $386 \pm 12$ & 0.9768 & 24.4 & 48.0 \\ \hline
    140 & 49.3 &  98.4 & 320 & $364 \pm 11$ & 0.9812 & 23.0 & 46.4 \\ \hline
    150 & 56.5 &  98.7 & 293 & $340 \pm 10$ & 0.9899 & 20.6 & 44.6 \\ \hline
    160 & 64.0 &  99.3 & 258 & $303 \pm  9$ & 0.9915 & 19.0 & 41.7 \\ \hline
    170 & 71.6 & 100.1 & 152 & $182 \pm  5$ & 0.9844 & 15.4 & 31.5 \\ \hline
    180 & 78.8 & 101.9 &  15 & $15.3\pm 0.5$ & 0.5615 &  9.3 & 10.0 \\ \hline
    185 & 81.4 & 104.0 &   7 & $6.2 \pm 0.2$ & 0.4213 &  7.6 &  7.1 \\ \hline
  \end{tabular}
  \caption{
Results for $\eetoXY$, $\XtoYg$, with $\myzero$.  Shown are values
for the mass $\mx$, the minimum and maximum energies 
($E_{min}$, $E_{max}$) allowed by the kinematic consistency cuts,
the number of observed events ($\rm N_{obs}$), the number of
events expected from $\eetonnggbra$ ($\rm N_{expected}$),
the value of $P_{\rm fluct}$ as defined in 
Equation~\protect\ref{eq:pfluct}, the calculated 95\% CL upper
limit on the number of new-physics signal events ($N^{95}$), and
the average value of $N^{95}$ expected in the absence of signal
($\langle N^{95}\rangle$).  Upper limits are calculated using the
event-counting method. Masses and energies are in GeV.
  }
\end{center}
\label{tab:spmy0}
\end{table}

\begin{table}[b]
\begin{center}
  \begin{tabular}{|c|c|c|c|c|} \hline

   $\delta$ & $\epsilon (\%)$ & $\langle \rm \sigma_{exp}^{95}\rangle$ (fb) & 
        $\sigma^{95}$ (fb)  & $M_D$ lower limit (GeV) \\
   \hline
   \hline
     2  & 76.5 & 241 & 309 & 1086 \\ \hline
     3  & 79.1 & 233 & 298 &  862 \\ \hline
     4  & 81.4 & 226 & 290 &  710 \\ \hline
     5  & 83.5 & 220 & 283 &  605 \\ \hline
     6  & 85.4 & 216 & 276 &  528 \\ \hline
     7  & 87.2 & 211 & 271 &  470 \\ \hline
  \end{tabular}
  \caption{
        Results for graviton-photon production in the context of extra
        dimensions.  Shown are the number of extra dimensions $\delta$,
        the efficiency $\epsilon$, 
        the average upper limit on the signal cross-section
        $\rm \langle \sigma_{exp}^{95} \rangle$ expected in the 
        absence of signal,
        the measured 95\% CL upper limit on signal cross-section $\sigma^{95}$,
        and the corresponding lower limit on the mass scale $M_D$.  
        Efficiencies and cross-section limits are evaluated
        within the restricted kinematic 
        acceptance described in Section~\ref{sec:sp_results_Gg}.
  }
\end{center}
\label{tab:sp_Gg}
\end{table}

\begin{table}
\centering
\begin{tabular}{|c||c|c|c|c|c|}
\hline
$\mx$ & $\my$=0 & $\my=\mx /2$ & $\my=\mx-10$ & $\my=\mx-5$ & $\my=\mx-2.5$ \\ \hline \hline
94 & $70.3\pm{1.2}$ & $70.7\pm{1.2}$ & $64.1\pm{1.3}$ & $38.9\pm{1.5}$ & $5.1\pm{0.7}$ \\ \hline
90 & $69.4\pm{1.2}$ & $71.9\pm{1.2}$ & $60.4\pm{1.4}$ & $40.5\pm{1.5}$ & $4.6\pm{0.7}$ \\ \hline
80 & $71.5\pm{1.2}$ & $69.3\pm{1.2}$ & $59.7\pm{1.4}$ & $39.5\pm{1.5}$ & $4.7\pm{0.7}$ \\ \hline
70 & $71.0\pm{1.2}$ & $70.8\pm{1.2}$ & $60.4\pm{1.4}$ & $44.0\pm{1.5}$ & $4.4\pm{0.7}$ \\ \hline
60 & $73.5\pm{1.1}$ & $71.1\pm{1.2}$ & $64.4\pm{1.3}$ & $42.5\pm{1.5}$ & $8.4\pm{0.9}$ \\ \hline
50 & $69.8\pm{1.2}$ & $69.2\pm{1.2}$ & $65.1\pm{1.3}$ & $45.4\pm{1.5}$ & $10.0\pm{0.9}$ \\ \hline
\end{tabular}
\caption[]{Acoplanar-photons selection efficiencies (\%) 
for the process $\eetoXX$, $\XtoYg$ at $\roots = 189$ GeV for various
$\mx$ and $\my$ (in GeV), after application of kinematic-consistency cuts.
These efficiency values are used to perform the efficiency
parameterization across the ($\mx$, $\my$) plane.  
The errors shown are due to Monte Carlo statistics only.
Efficiencies for the generated points at $\my = 20$ and $\my=\mx-15$
are not shown, but are similar to those for $\my = 0$ and $\my = \mx/2$.
}
\label{tab:g2_eff_189}
\end{table}

\begin{table}
\centering
\begin{tabular}{|c||c|c|c|c|}
\hline
 & Selection efficiency & Selection efficiency with & $\rm N_{data}$ & 
$\rm N_{\nngggbra}$ \\
$\mx$ &  & $\mxmax>\mx-5$ GeV &  & \\ \hline
\hline
94 & $72.2 \pm 1.2$  & $70.4 \pm 1.2$ & 3 & $1.34\pm{0.07}$ \\ \hline
90 & $71.3 \pm 1.2$  & $67.5 \pm 1.3$ & 5 & $2.40\pm{0.09}$ \\ \hline
80 & $72.3 \pm 1.2$  & $68.7 \pm 1.4$ & 7 & $4.81\pm{0.13}$ \\ \hline
70 & $71.4 \pm 1.2$  & $69.2 \pm 1.2$ & 9 & $7.22\pm{0.15}$ \\ \hline
60 & $74.0 \pm 1.1$  & $71.1 \pm 1.2$ & 11 & $10.05\pm{0.18}$ \\ \hline
50 & $70.2 \pm 1.2$  & $67.7 \pm 1.3$ & 14 & $13.67\pm{0.20}$ \\ \hline
\end{tabular}
\caption[]{
Acoplanar-photons event selection efficiencies (\%), as a function of mass, 
for the process $\eetoXX$, $\XtoYg$, for $\myzero$ at 
$\roots = 189$ GeV. The first column shows the efficiency  
of the selection described in section~3.2.
The second column shows the efficiency (\%) after the 
additional cut on $\mxmax$.  The third column
shows the number of selected events consistent with the mass value $\mx$.
The last column shows the expected number of events from the process 
$\eetonngggbra$ (KORALZ).  The errors shown due to from Monte Carlo 
statistics only. 
}
\label{tab:g2_eff_my0_189}
\end{table}


\begin{table}
\centering
\begin{tabular}{|c||c|c|c|c|}
\hline
                    & \multicolumn{4}{c|}{Selection efficiency with $\mxmax>\mx-5$ GeV} \\
\hline
 $\log_{10}(\tau_X)$  & $\mx$=50 GeV &  $\mx$=75 GeV &  $\mx$=90 GeV &  $\mx$=94 GeV \\
\hline
-15 & 65.2 $ \pm $  1.7 & 68.2 $ \pm $  1.6 & 71.1 $ \pm $  1.5 & 72.1 $ \pm $  1.5 \\ \hline
-10 & 63.3 $ \pm $  1.8 & 67.8 $ \pm $  1.6 & 72.2 $ \pm $  1.5 & 71.4 $ \pm $  1.5 \\ \hline
 -9 & 59.9 $ \pm $  1.8 & 71.7 $ \pm $  1.6 & 71.3 $ \pm $  1.6 & 72.1 $ \pm $  1.6 \\ \hline
-8.3& 25.8 $ \pm $  1.7 & 43.5 $ \pm $  4.0 & 49.7 $ \pm $  4.8 & 51.4 $ \pm $  5.6 \\ \hline
-8  & 13.9 $ \pm $  1.5 & 25.8 $ \pm $  3.0 & 33.0 $ \pm $  3.8 & 37.8 $ \pm $  3.7 \\ \hline
-7.3&  2.5 $ \pm $  1.1 &  2.7 $ \pm $  1.0 &  5.6 $ \pm $  1.1 & 10.7 $ \pm $  2.4 \\ \hline
-7  &  0.8 $ \pm $  1.0 &  1.0 $ \pm $  0.9 &  2.2 $ \pm $  1.0 &  5.1 $ \pm $  1.2 \\ \hline
\hline
${\rm N_{data}}$   &  14 &  7  &  5  & 3 \\ \hline
$\rm N_{\nngggbra}$  & 13.67  $ \pm $ 0.20 & 5.88   $ \pm $ 0.14 & 2.40   $ \pm $ 0.09 & 1.34   $ \pm $ 0.07 \\ \hline
\end{tabular}
\caption[]{
Acoplanar-photons event selection efficiencies (\%), 
as a function of mass $\mx$, and lifetime $\tau_X$
for the process $\eetoXX$, $\XtoYg$, for $\myzero$ at 
$\roots = 189$ GeV. The uncertainties in the selection efficiencies 
include the contributions due to the limited Monte Carlo signal sample size,
the variation of the timing cuts, and the parameterisation      
of the fit efficiencies. Also included are 
the number of events found in the data and the number 
expected from the $\eetonngggbra$ background that are 
consistent with the kinematic consistency cut for the 
$\mx$ mass values considered. 
}
\label{tab:g2_eff_my0_189_np}
\end{table}


%
%
%
\newpage
\begin{figure}[b]
        \centerline{\epsffile{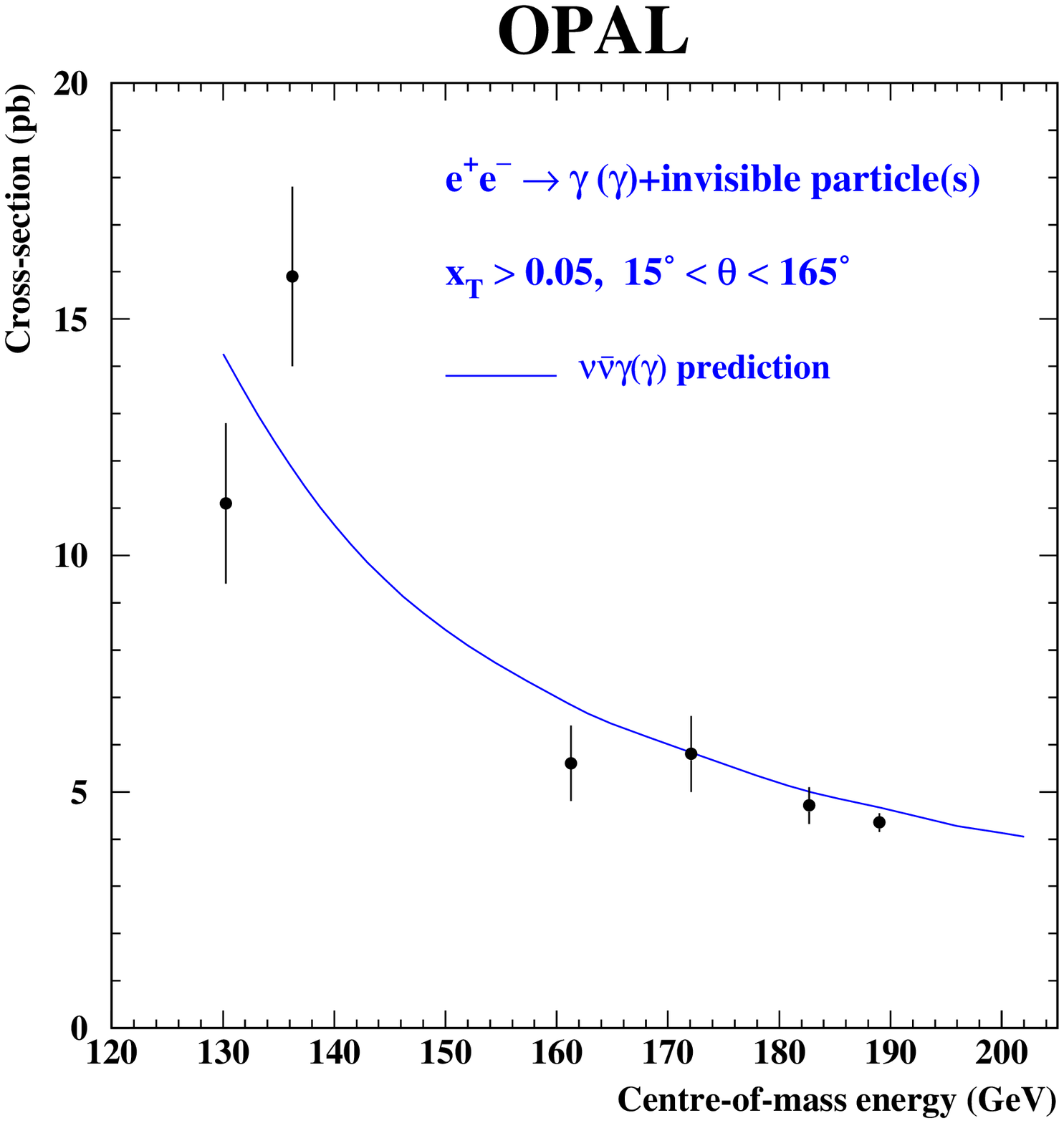}}

        \caption{ The measured value of 
        $\sigma(\epem \to \gamma (\gamma)$ + invisible particle(s)),
        within the kinematic acceptance of the single-photon selection,
        as a function of $\roots$.
        The data points with error bars are OPAL
        measurements at $\roots$ = 130, 136, 161, 172, 183 and 189~GeV. 
        The curve is the prediction for the
        Standard Model process $\eetonnggbra$ from the KORALZ generator.
        }
\label{f:sp_xs130_189}
\end{figure}
%
\newpage
\begin{figure}[b]
\centerline{\epsffile{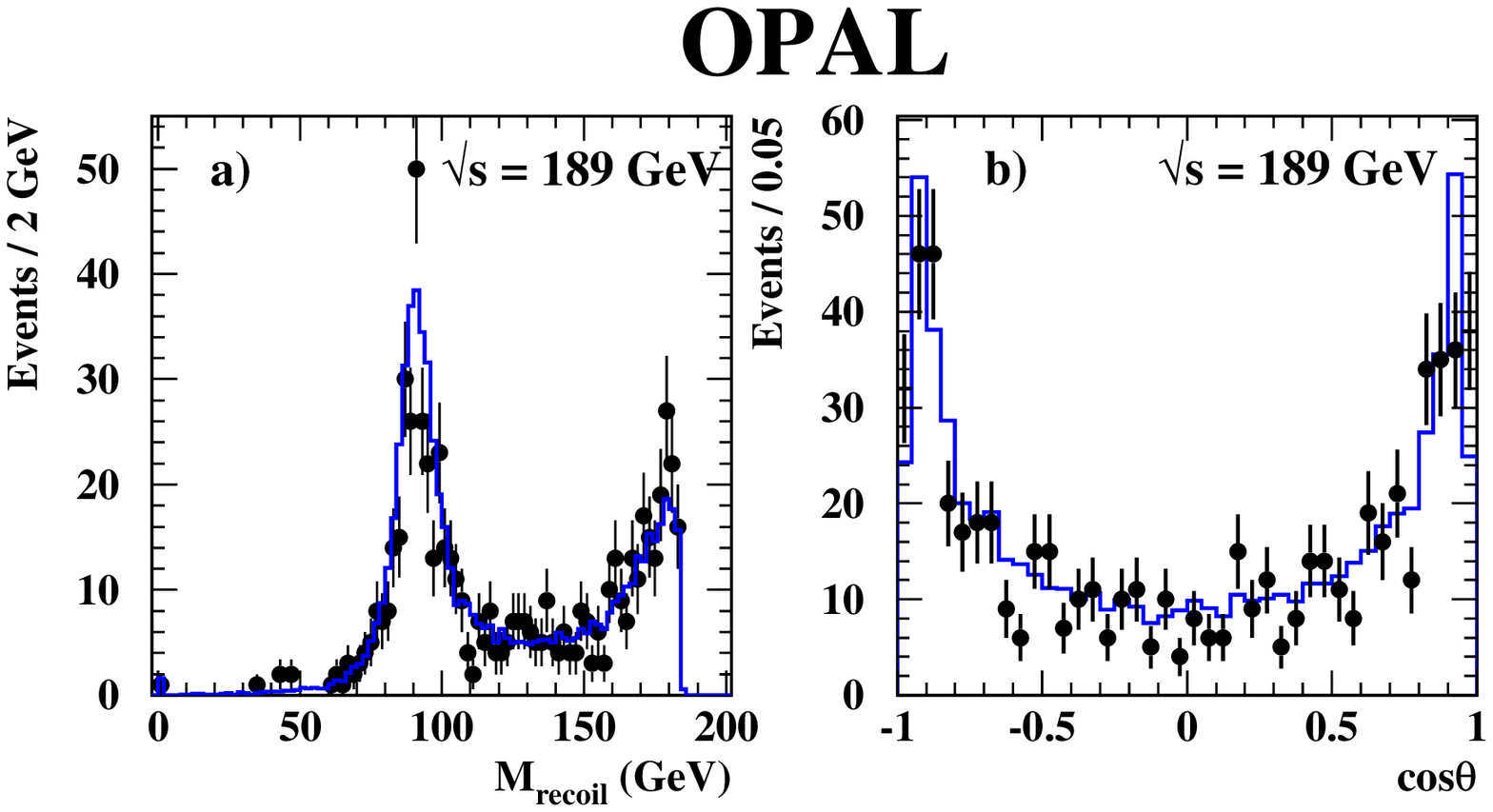}}
\caption{a)~The recoil mass distribution for events passing the
single-photon selection for the $\roots$ = 189~GeV data sample.
b)~The $\cos\theta$ distribution for 
the most energetic photon in the single-photon selection
at $\roots$ = 189 GeV. 
In both plots, the points with error bars are the data and the 
histogram is the expectation from the KORALZ $\eetonnggbra$ Monte Carlo
normalized to the integrated luminosity of the data.
} 
\label{f:sp_data189}
\end{figure}
%
\clearpage
\newpage
\begin{figure}[b]
\centerline{\epsfig{file=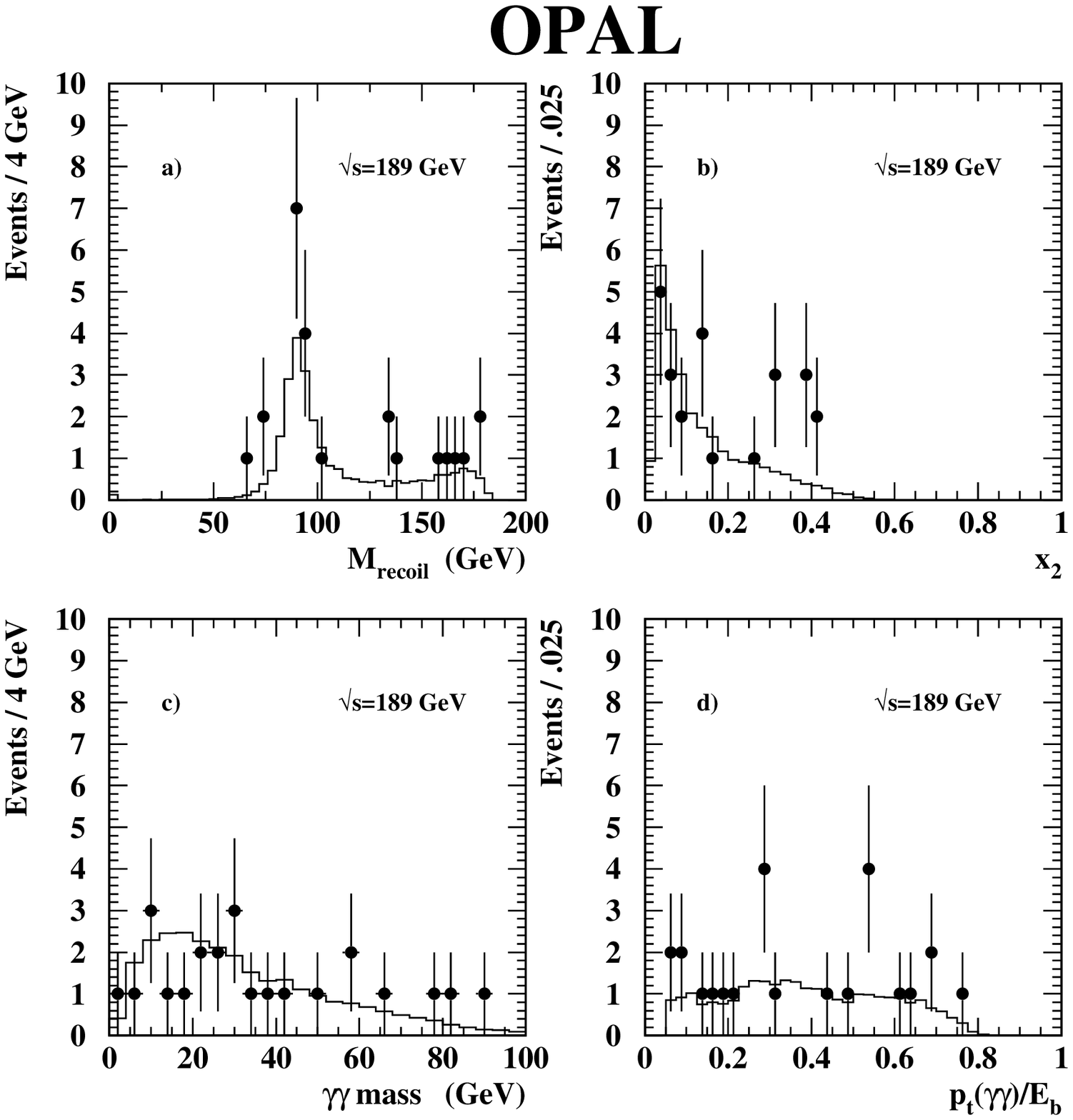,
width=16cm,bbllx=20pt,bblly=140pt,bburx=540pt,bbury=700pt}}
\caption{Plots of kinematic quantities for the
selected acoplanar-photons events for
$\sqrt{s}=$ 189 GeV.
a) Recoil-mass distribution. b) Distribution of the scaled energy
of the second photon (x$_2$). c) Distribution of the invariant mass
of the $\gamgam$ system. d) Scaled transverse momentum distribution for
the $\gamgam$ system.
The data points with error bars represent the selected OPAL data events.
In each case the histogram shows the expected contribution from 
$\eetonngggbra$ events, from KORALZ, normalized to the integrated luminosity
of the data. 
}
\label{f:g2_kine_189}
\end{figure}
\newpage
\begin{figure}[b]
        \centerline{\epsffile{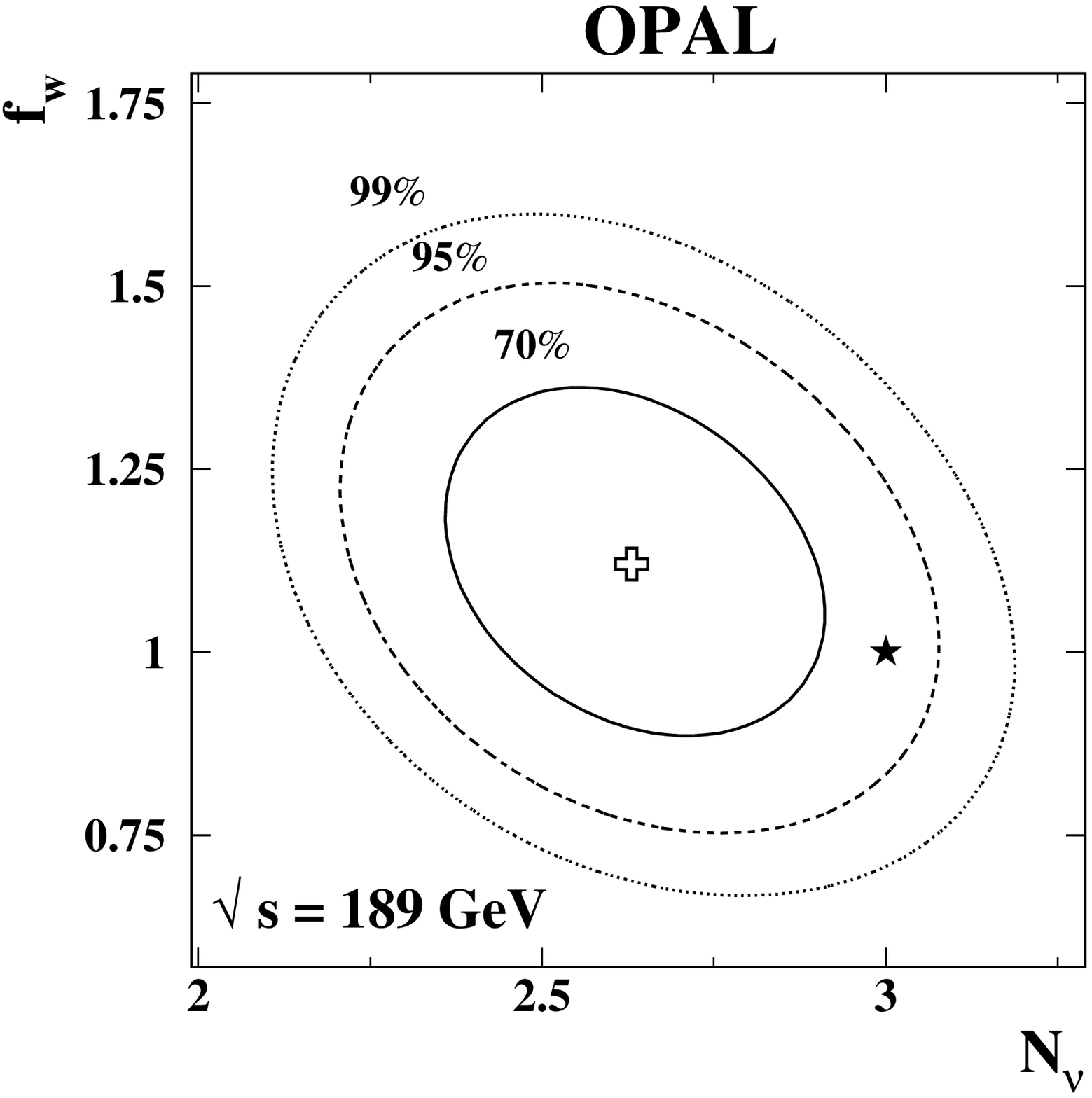}}
        \caption{ The 70\%, 95\% and 99\% confidence 
        level contours in the $f_W$,$\rm{N_{\nu}}$ plane
        resulting from the likelihood fit
        to the overall rate and the photon energy spectrum 
        measured in the single-photon selection. The cross 
        indicates the central values of the fit results, while  
        the Standard Model expectation ($N_\nu$=3, $f_W$=1) is shown by 
        the star.
        }
\label{f:sp_nnucont}
\end{figure}
\newpage
\begin{figure}[b]
        \centerline{\epsffile{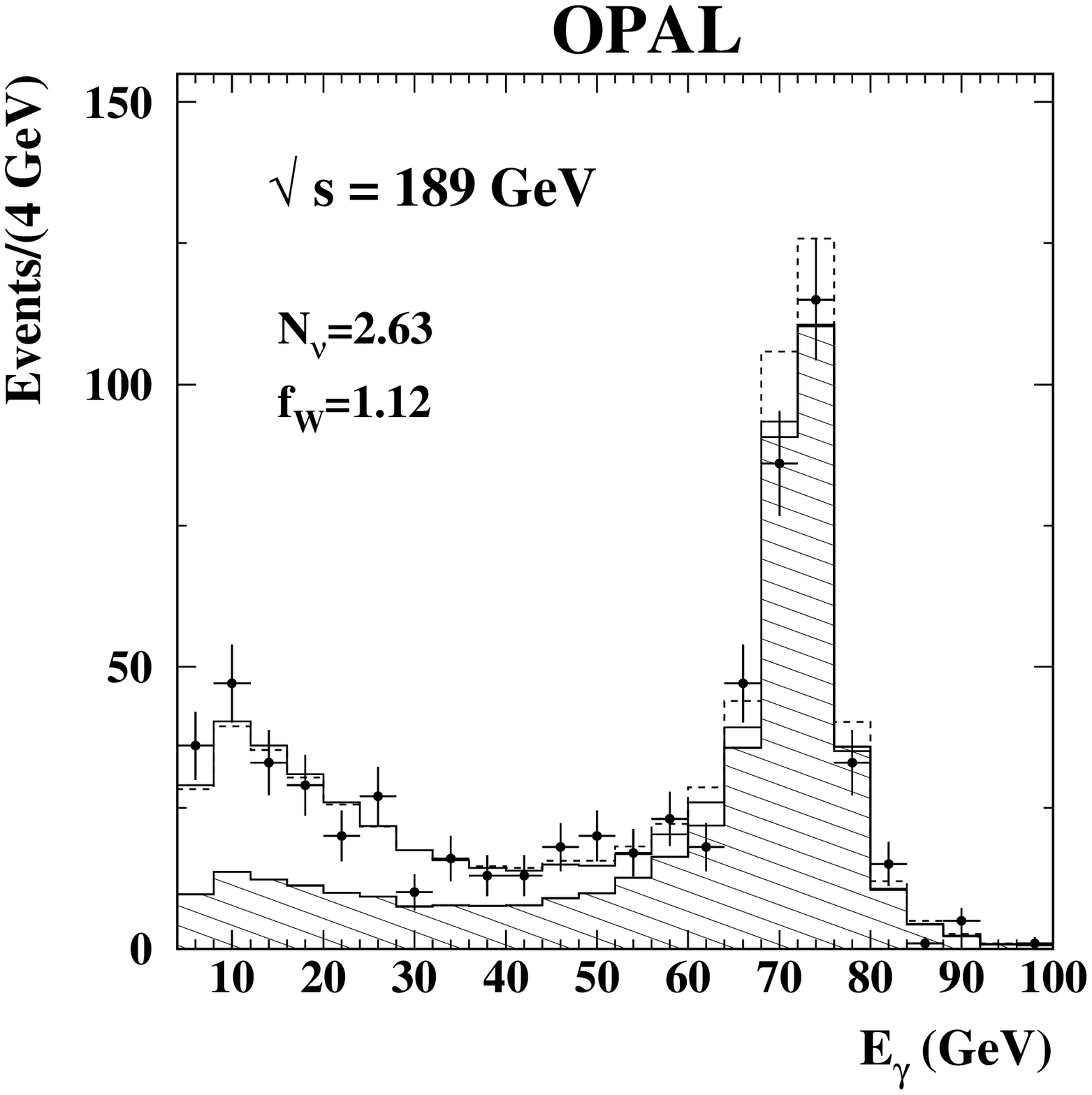}}

        \caption{Photon energy distributions for single-photon events.  
        The points with error bars are the data.  The solid histogram is 
        the prediction for the values $f_W = 1.12$, $\rm N_{\nu}=2.63$
        most consistent with the data.  The dashed
        histogram is the expectation for the Standard Model values 
        $f_W = 1$, $\rm N_{\nu}=3$.  The hatched region indicates the 
        pure s-channel $\rm Z^{0}$ contribution for $\rm N_{\nu}=2.63$.  
        All predicted distributions were calculated using the NUNUGPV98 
        generator.}

\label{f:sp_nnueg}
\end{figure}
%
%
\newpage
\begin{figure}[ht]
\centerline{\epsfig{file=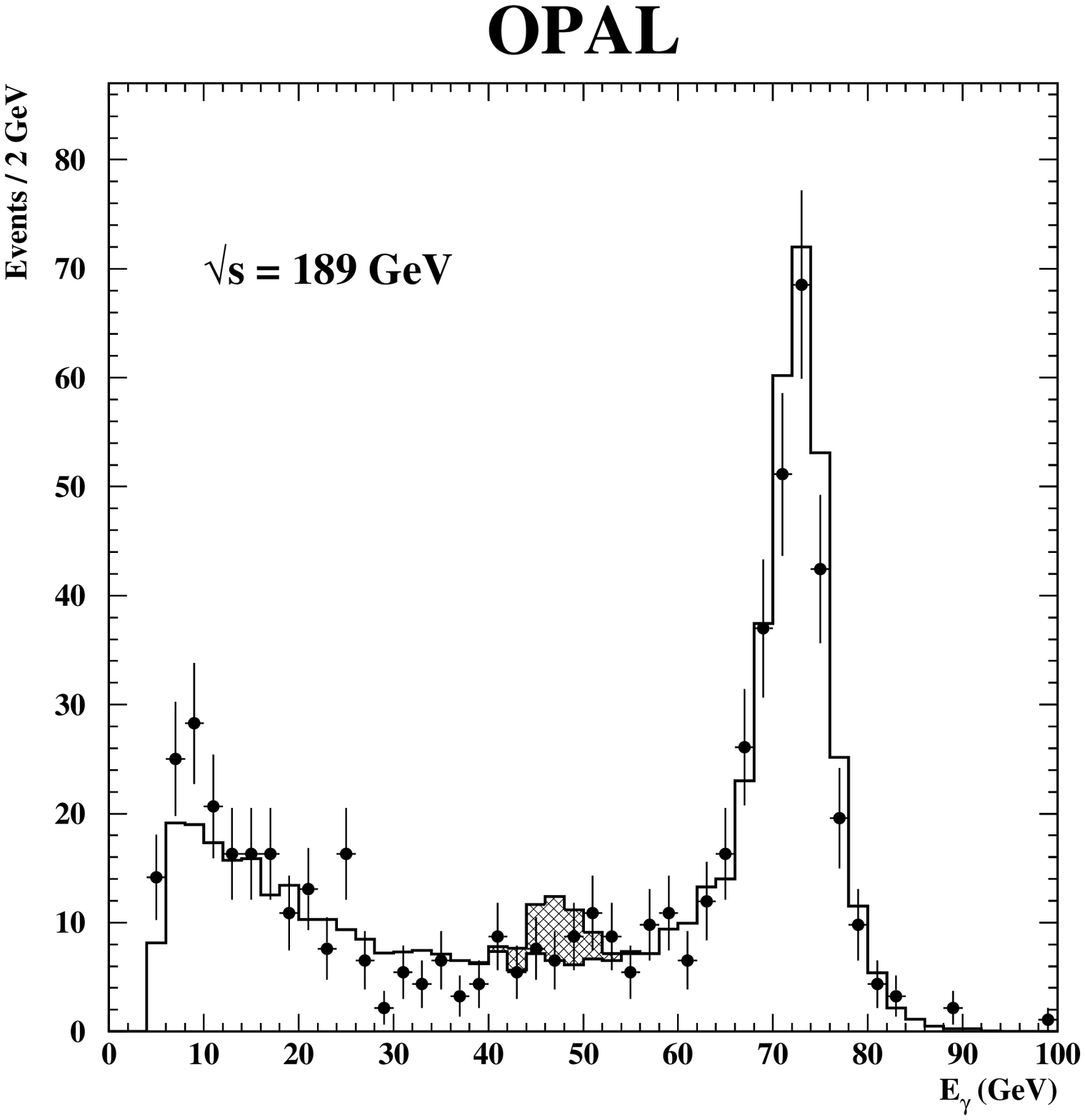,
        width=16cm,bbllx=10pt,bblly=140pt,bburx=530pt,bbury=650pt}}
        \caption{The observed single-photon energy distribution after 
            application of the degraded-resolution cuts.  The data 
            are shown as the points with error bars and are compared 
            with the expectation from the Standard Model process
            $\eetonnggbra$, evaluated with KORALZ.
            The extra hatched contribution around 
            45 GeV illustrates the expected additional contribution 
            from XY production which is excluded at 95\% CL using 
            the event-counting method for particular X and Y masses.
            The hatched contribution corresponds to $\mx$=125~GeV and 
            $\my$=63~GeV with a cross-section times branching ratio 
            of 142 fb. 
        }
        \label{f:sp_likeresult}
\end{figure}
%
%
%
\newpage
\begin{figure}[ht]
        \centerline{\epsffile{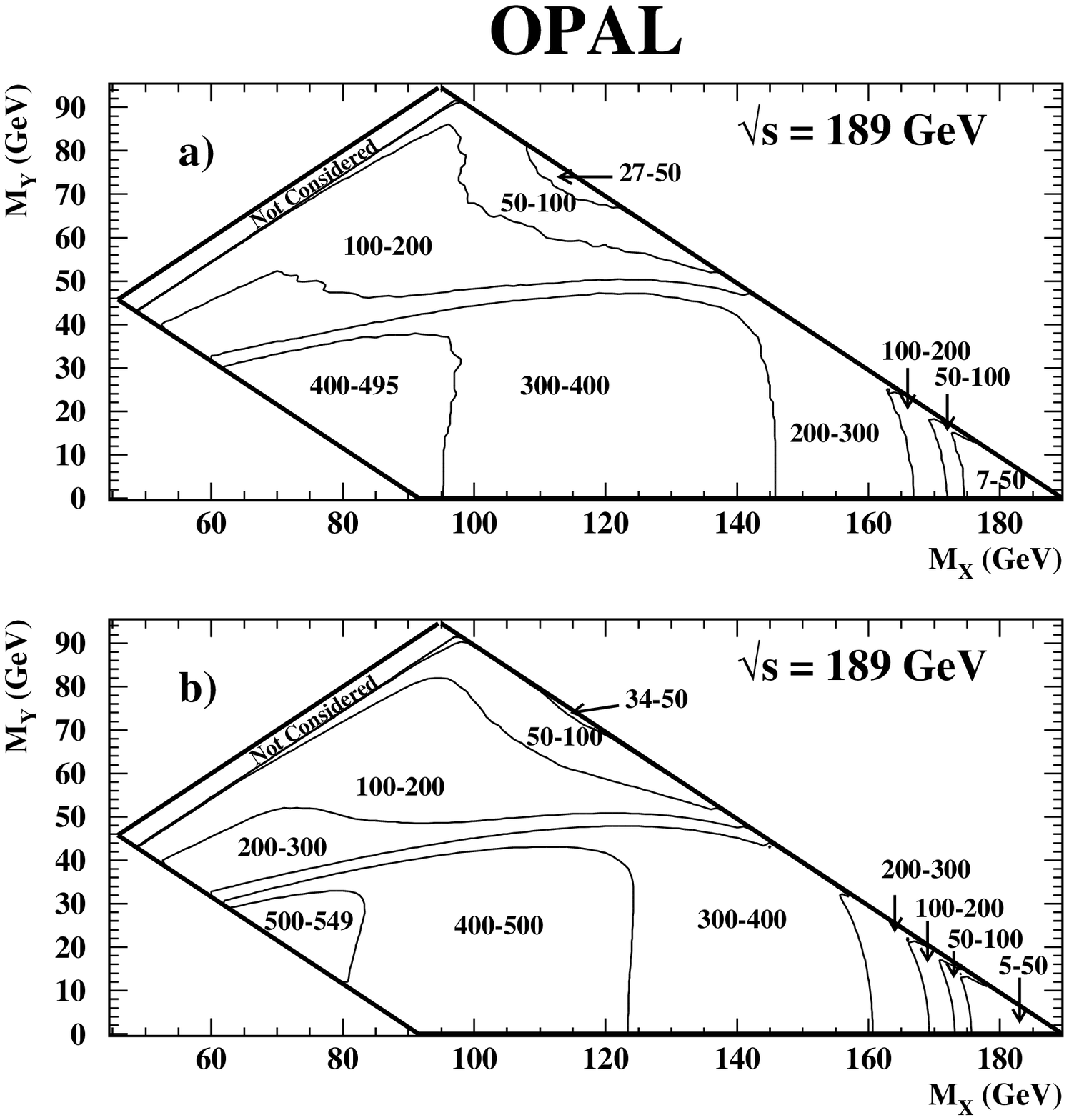}}
        \caption{ 
                Contour plots for the $\sqrt{s} =$ 189 GeV data sample
                search for $\eetoXY$, $\XtoYg$, after application of
                both degraded-resolution and kinematic-consistency cuts.
                For each set of 
                mass values ($\mx$, $\my$), a) shows the range of 
                numbers of observed
                single-photon candidate events between each contour
                and b) shows the range of numbers of expected events
                from $\eetonnggbra$ between each contour.
                The $\nnggbra$ expectation was derived using KORALZ.
                Lines are drawn around the boundaries defined by 
                $\mx + \my = 189$ GeV, $\mx = \my$, and 
                $\mx + \my = M_{\rm Z}$.
                The region in which $\mx - \my < 5$~GeV is not 
                considered.}
        \label{f:sp_evsXY_data}
\end{figure}
%
%
%
\newpage
\begin{figure}[ht]
        \centerline{\epsffile{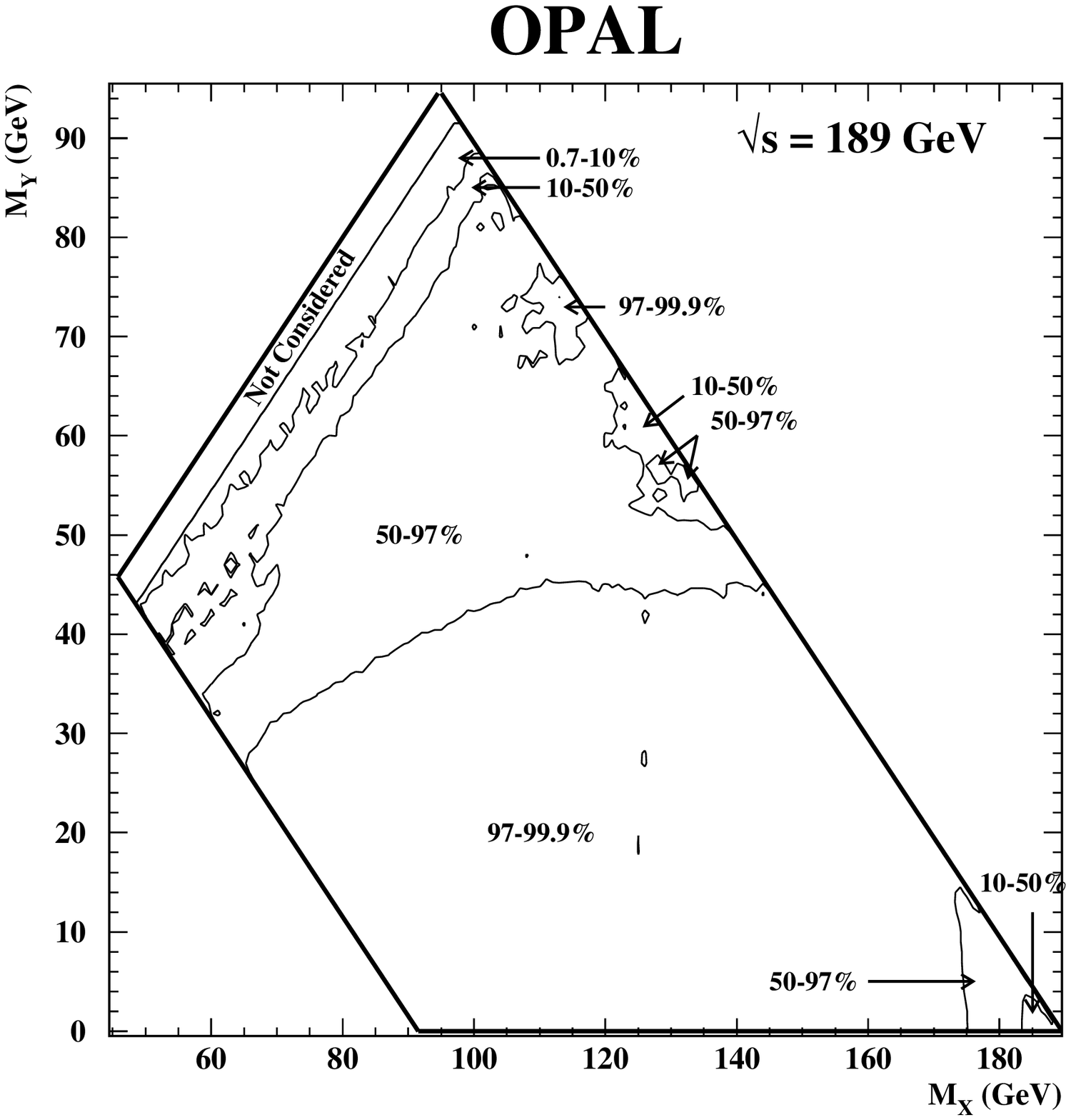}}
        \caption{Consistency between the number of observed events 
                and the number of events expected from $\eetonnggbra$
                as estimated by KORALZ.  Plotted is the probability
                for the number of expected events to fluctuate to
                the number of observed events or more, as described
                in the text.
                Values greater than 50\% indicate a downward fluctuation
                where the number of observed events is fewer than expected.
                The boundaries and delineated regions are as defined for 
                Figure~\ref{f:sp_evsXY_data}.}
        \label{f:sp_consistency}
\end{figure}
%
\newpage
\begin{figure}[ht]
        \centerline{\epsffile{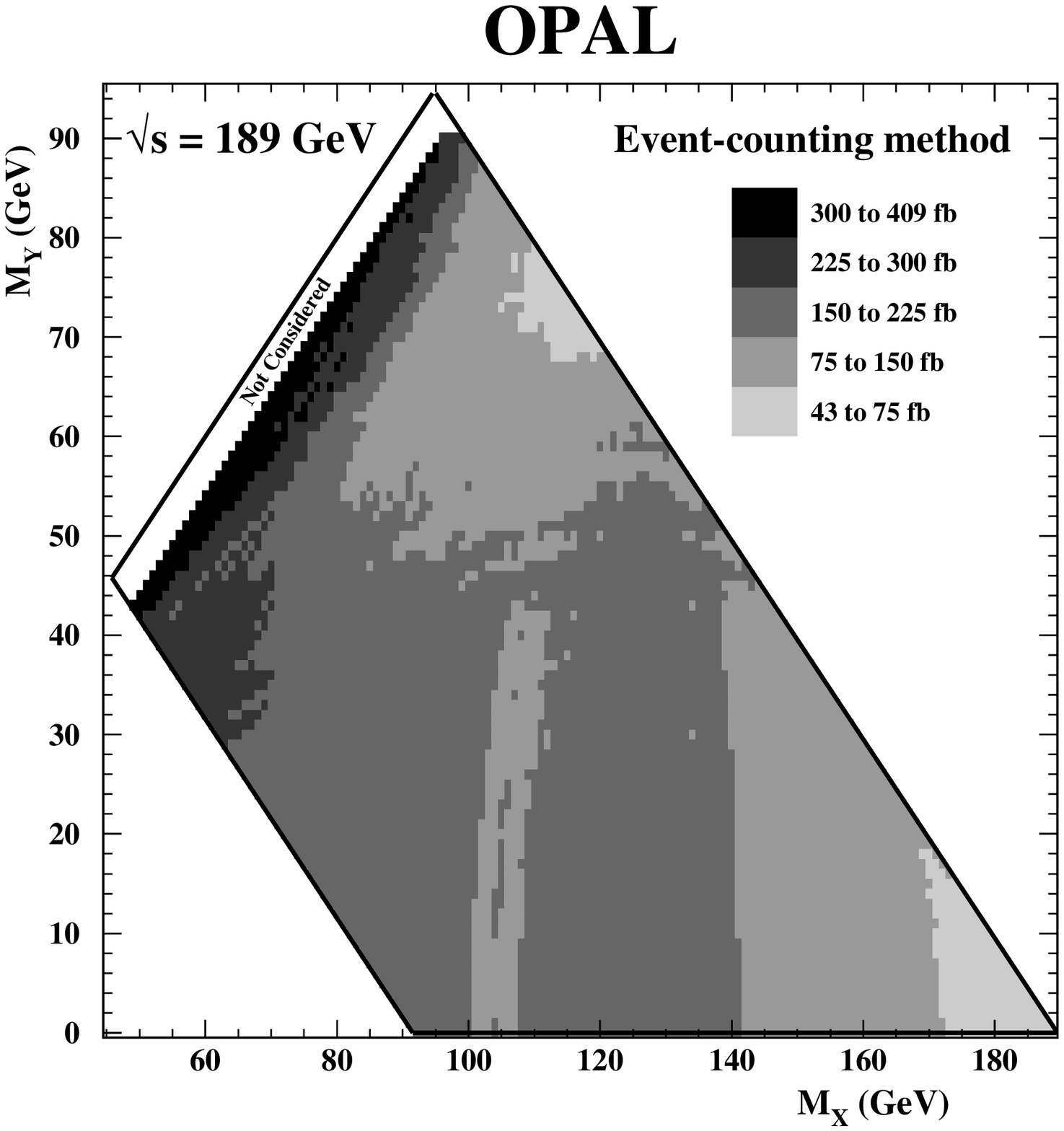}}
        \caption{The 95\% CL upper limit on $\sigbrXY$ 
                at $\roots = 189$ GeV
                as a function of $\mx$ and $\my$, using the 
                event-counting method described in the text.
                The boundaries and delineated regions are as defined for 
                Figure~\ref{f:sp_evsXY_data}.}
        \label{f:sp_countlim}
\end{figure}
%
%
%

%
\newpage
\begin{figure}[ht]
        \centerline{\epsffile{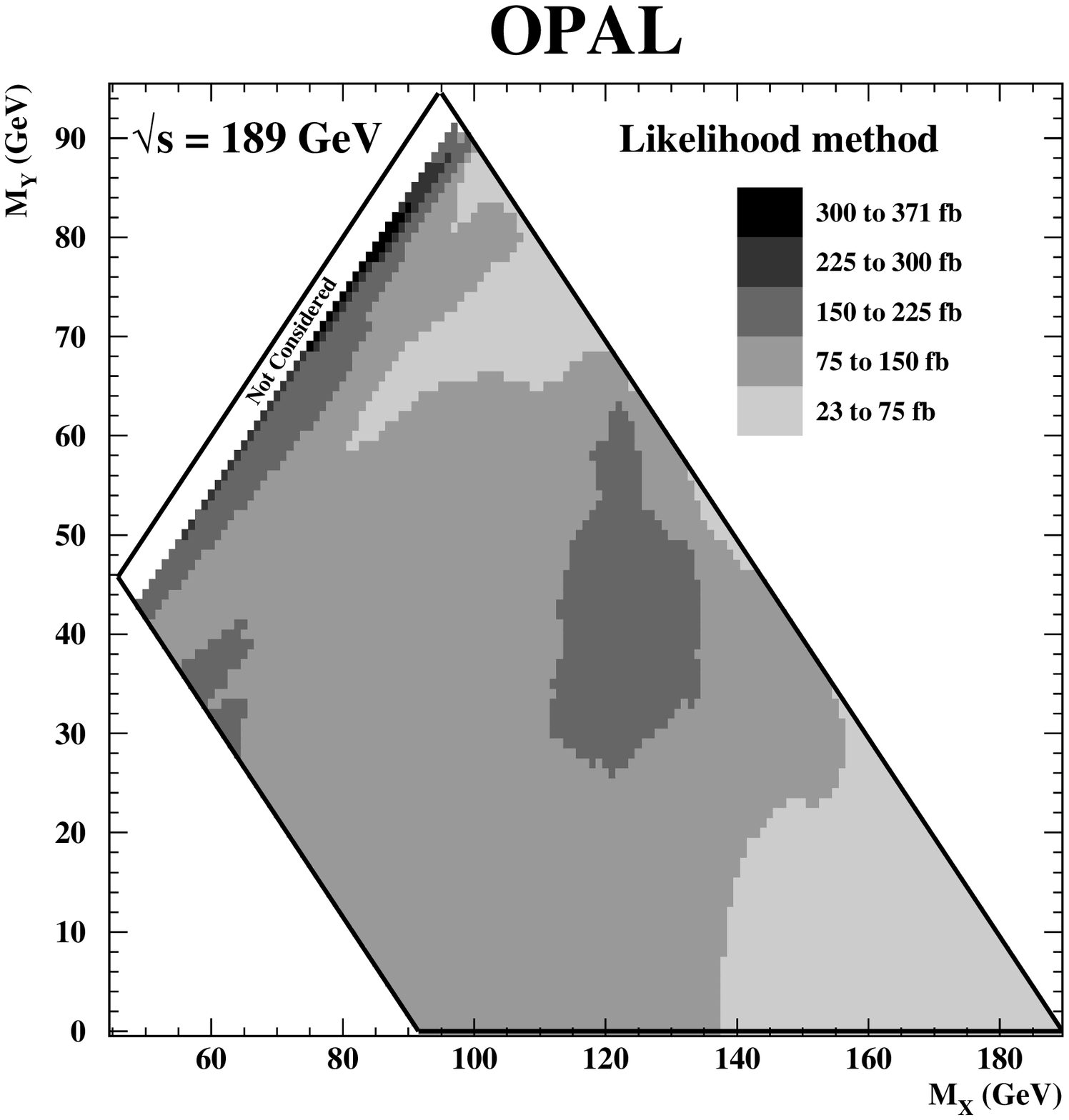}}
        \caption{The 95\% CL upper limit on $\sigbrXY$ 
                at $\roots = 189$ GeV
                as a function of $\mx$ and $\my$, using the likelihood-based
                method described in the text.
                The boundaries and delineated regions are as defined for 
                Figure~\ref{f:sp_evsXY_data}.}
        \label{f:sp_likelim}
\end{figure}
%
%
\newpage
\begin{figure}[ht]
        \centerline{\epsffile{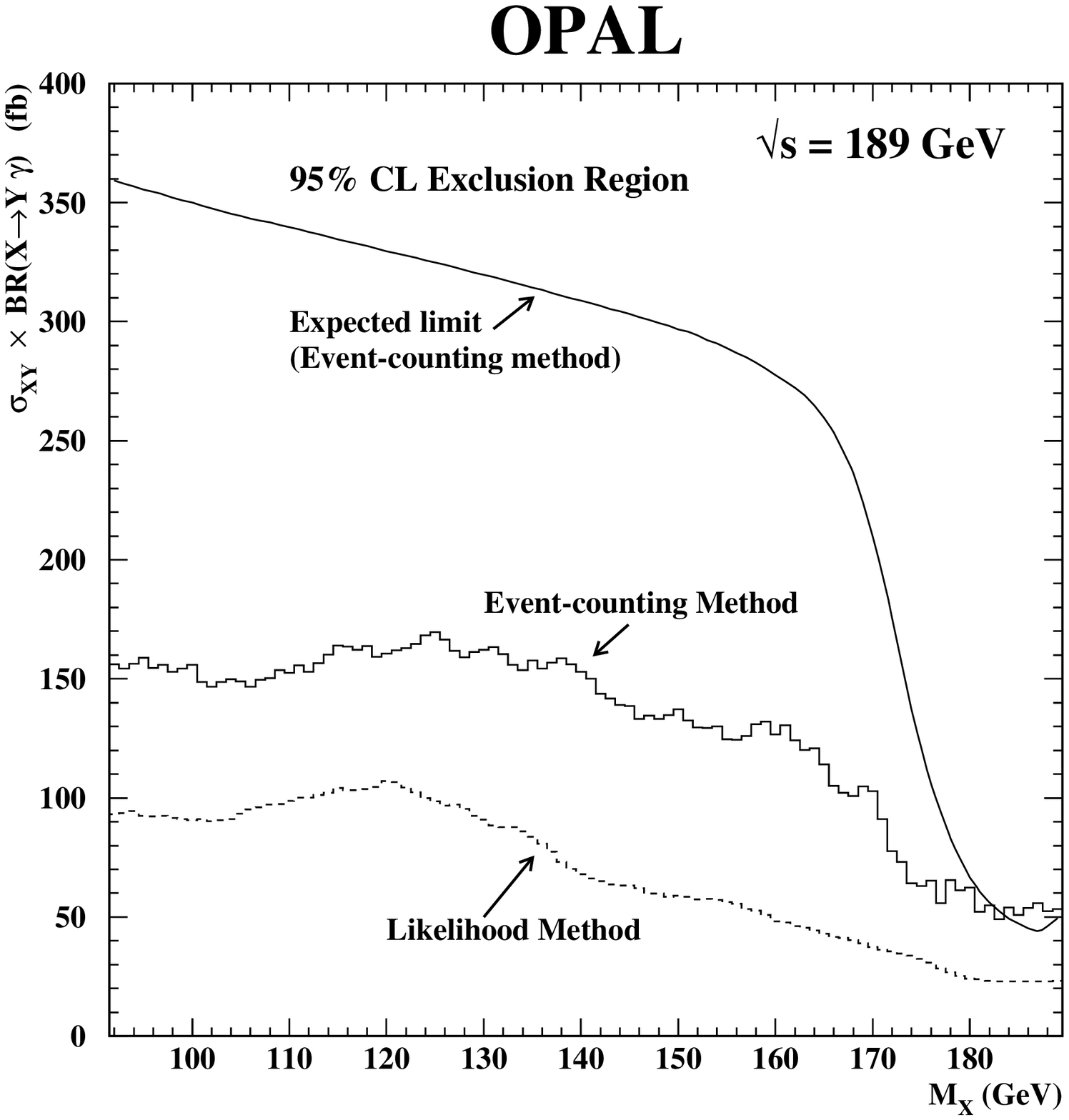}}
        \caption{The 95\% CL upper limits on $\sigbrXY$ at $\roots = 189$ GeV
                as a function of $\mx$, assuming $\myzero$, calculated 
                using the likelihood-based method and the event-counting
                method as described in the text.  Also shown is the
                average limit expected in the absence of signal, calculated
                using the event-counting method.}
        \label{f:sp_XYlim_massless}
\end{figure}
%
\newpage
\begin{figure}[b]
\centerline{\epsfig{file=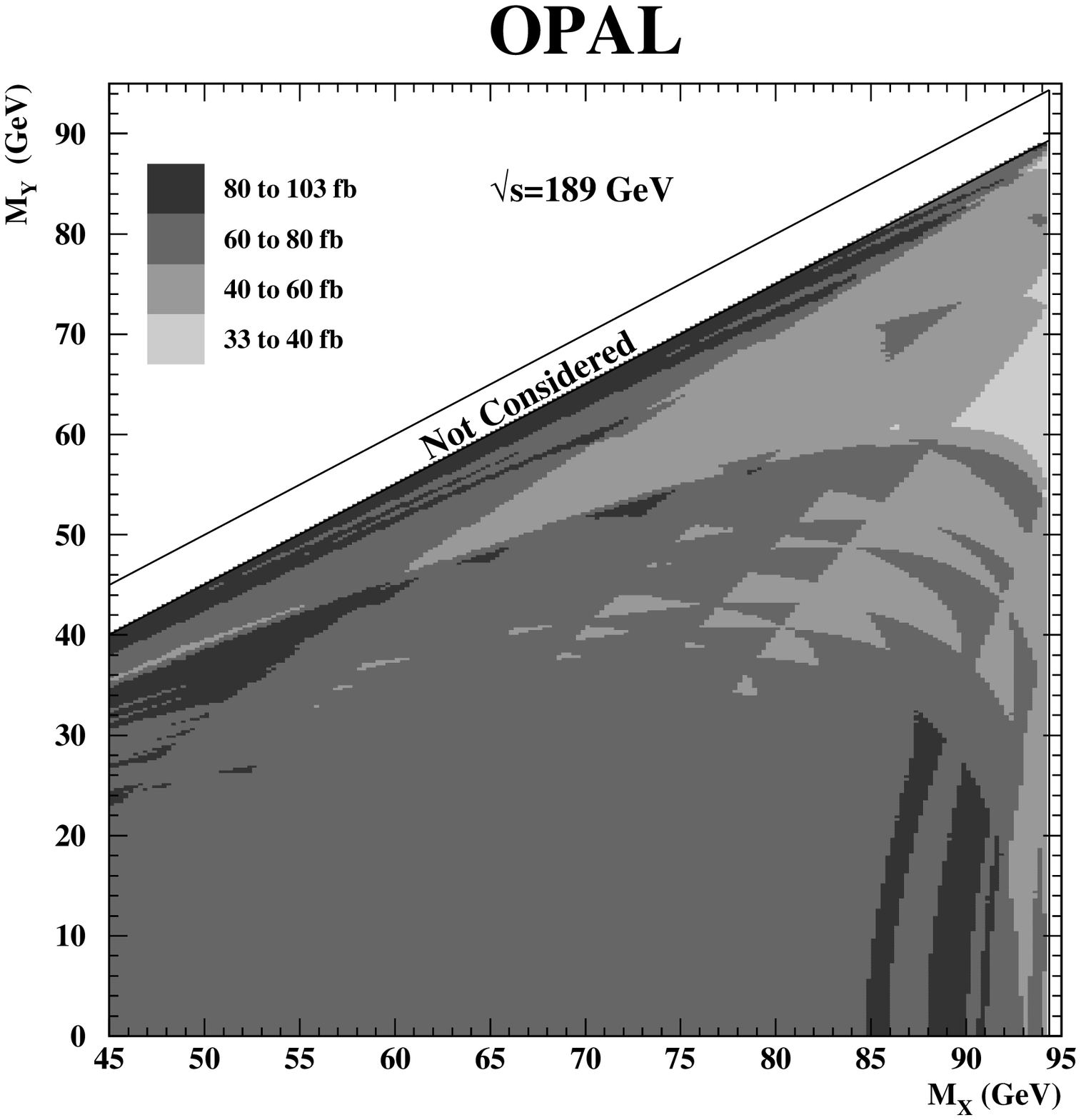,
width=16cm,bbllx=20pt,bblly=140pt,bburx=540pt,bbury=700pt}}
\caption{
The shaded areas show 95\% CL exclusion regions
for $\sigbrXX$  at $\roots = 189$ GeV. 
No limit is set for mass-difference values 
$\mx-\my < 5$ GeV, defined by the lower line above 
the shaded regions. The upper line is for $\mx=\my$.
}
\label{f:g2_mxmy_189}
\end{figure}
\newpage
\begin{figure}[b]
\centerline{\epsfig{file=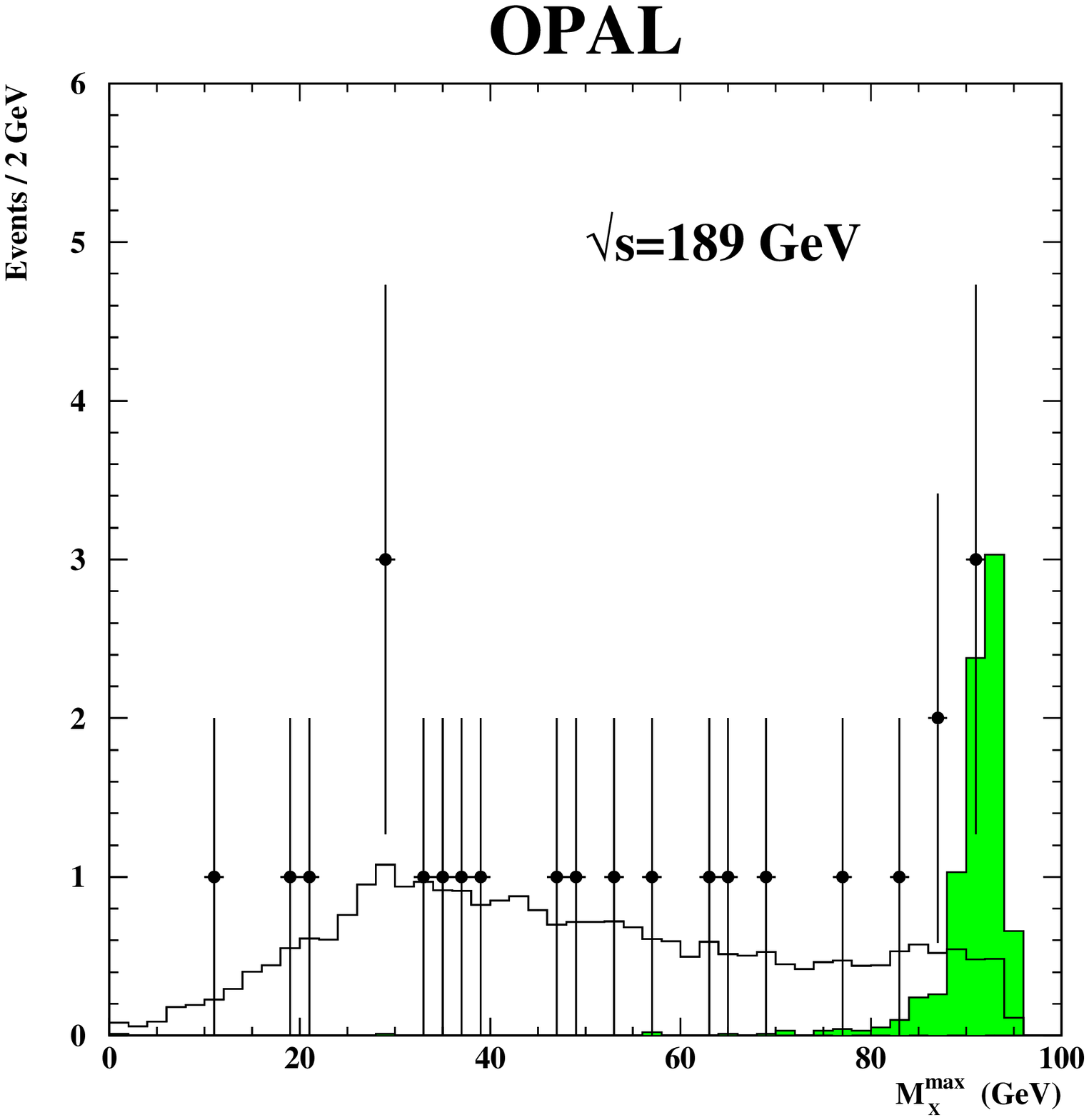,width=18cm,
bbllx=20pt,bblly=140pt,bburx=540pt,bbury=700pt}}
\caption{
$\mxmax$ distribution of the selected acoplanar-photons events
(points with error bars). Shown as an unshaded histogram is the expected 
distribution from the Standard Model process $\eetonngggbra$, 
evaluated using KORALZ and
normalized to the integrated luminosity of the data. The shaded histogram 
shows the expected distribution for the signal process $\eetoXX$, $\XtoYg$ for 
$\mx = 90$ GeV with arbitrary production cross-section.}
\label{f:g2_mxmax_189}
\end{figure}
\newpage
\begin{figure}[b]
\centerline{\epsfig{file=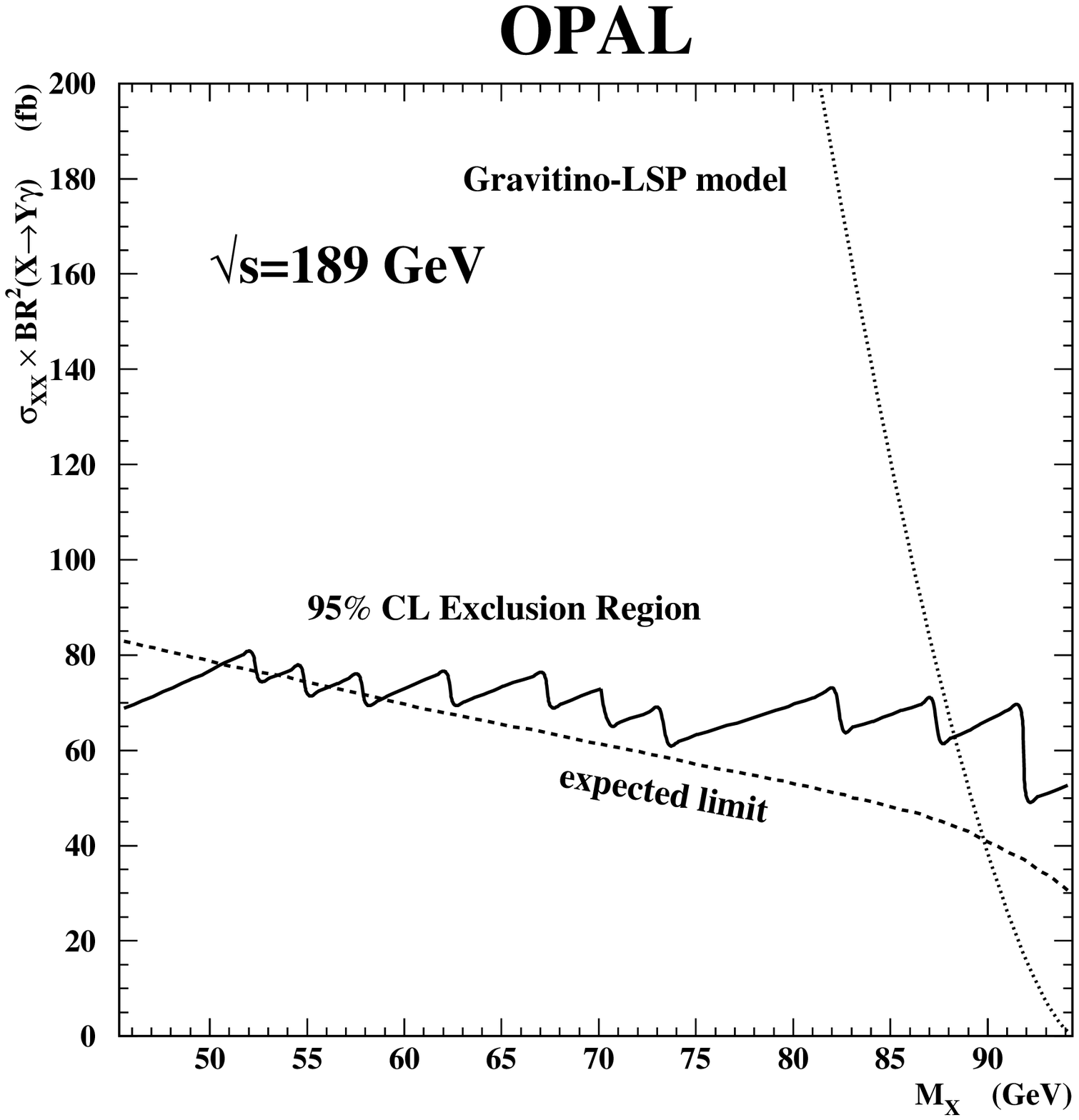,width=16cm,
bbllx=20pt,bblly=140pt,bburx=540pt,bbury=700pt}}
\caption{
95\% CL upper limit on $\sigbrXX$ 
for $\myzero$ (solid line). Also shown is the expected limit
(dashed line). The dotted line shows 
the cross-section prediction of
a specific light gravitino LSP model\cite{chang}. 
Within that model, $\lsp$ masses between 45 and 88.3 GeV are excluded 
at the 95\% CL. These limits assume that particle X decays promptly.}
\label{f:g2_limit_my0_189}
\end{figure}
\begin{figure}[b]
\centerline{\epsfig{file=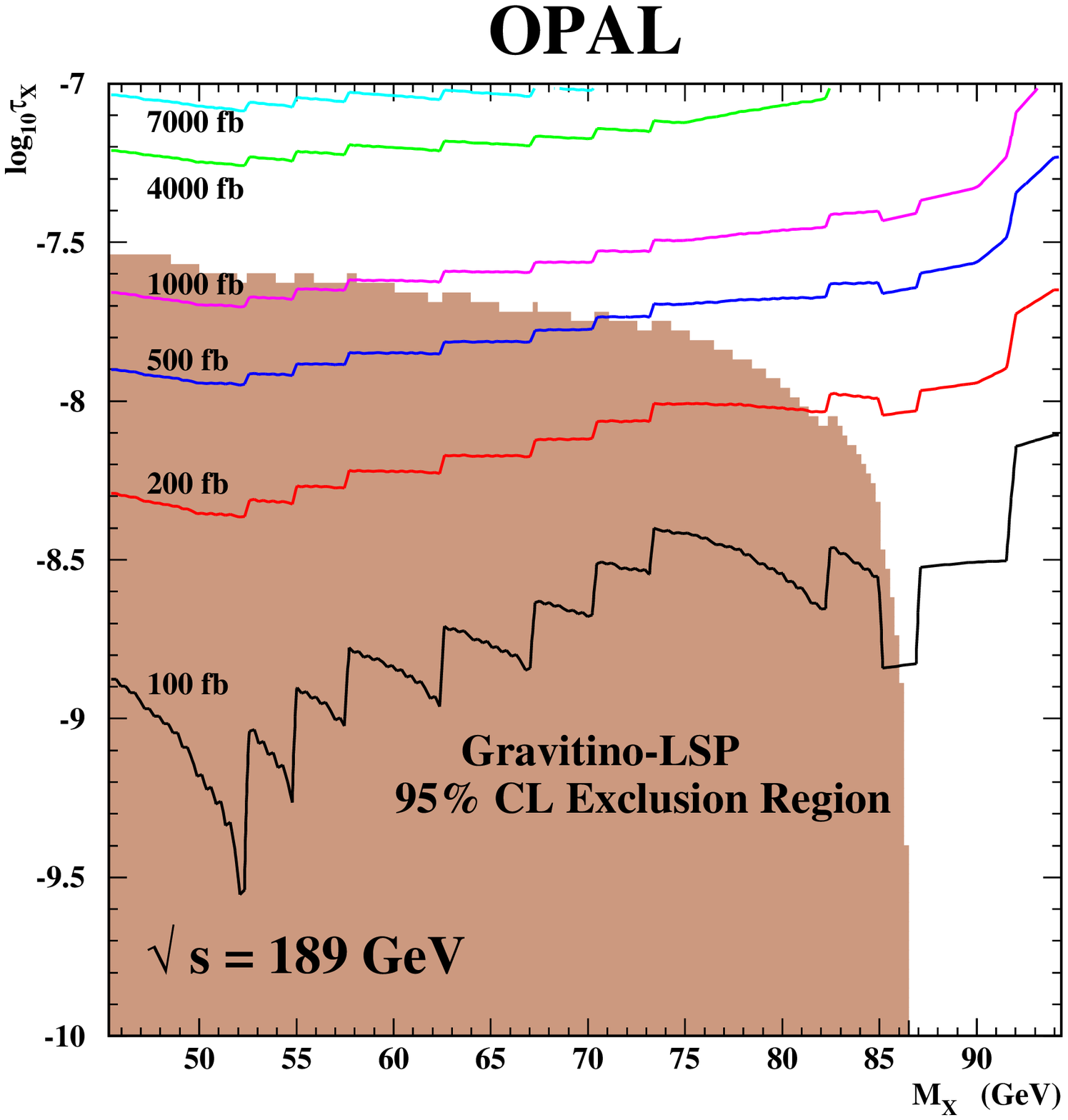,width=16cm,
bbllx=20pt,bblly=140pt,bburx=540pt,bbury=700pt}}
\caption{
The 95\% CL upper limit on $\sigbrXX$ 
for $\myzero$ as a function of $\mx$ and $\log_{10}(\tau_X)$ 
(with $\tau_X$ in seconds), calculated using the efficiencies of 
Table \ref{tab:g2_eff_my0_189_np} to extend the predictions of 
Figure \ref{f:g2_limit_my0_189} to the case of $\PX$ with a 
non-negligible decay length. Superimposed is the domain defined by the 95\% 
CL exclusion limit (shaded region) under the assumption of a light gravitino 
LSP model  \cite{chang}.
}
\label{f:g2_limit_mx_tau_pr189}
\end{figure}

\end{document}